\documentclass[twocolumn, tighten]{aastex631}
\usepackage{natbib}
\usepackage{xspace}

\newcommand{\upenn}{Department of Physics \& Astronomy, University of Pennsylvania, 209 S 33rd St., Philadelphia, PA 19104, USA}
\newcommand{\cca}{Center for Computational Astrophysics, Flatiron Institute, 162 5th Ave, New York, NY 10010, USA}
\newcommand{\yale}{Department of Astronomy and Astrophysics, Yale University, New Haven, CT 06520, USA}
\newcommand{\ifa}{Institute for Astronomy, University of Hawai`i, 2680 Woodlawn Drive, Honolulu, HI 96822, USA}
\newcommand{\columbia}{Department of Astronomy, Columbia University, Pupin Physics Laboratories, New York, NY 10027, USA}
\newcommand{\Msun}{\ensuremath{M_{\odot}}}

\newcommand{\degree}{\ensuremath{{}^{\circ}}\xspace}
\newcommand{\degrees}{\degree}
\newcommand{\mycomment}[1]{}
\newcommand{\code}[1]{\texttt{#1}\xspace}

\newcommand{\COMMENT}[3]{\textcolor{#1}{#2: #3}}
\newcommand{\NP}[1]{\COMMENT{magenta}{Nondh}{#1}}

\graphicspath{{./}{figures/}}

\shorttitle{Constraining the Tilt of DM Halo}
\shortauthors{Panithanpaisal et al.}

\begin{document}

\title{Constraining the Tilt of the Milky Way's Dark Matter Halo with the Sagittarius Stream}

\correspondingauthor{Nondh Panithanpaisal}
\email{nondh@sas.upenn.edu}

\author[0000-0001-5214-8822]{Nondh Panithanpaisal}
\affil{\upenn}

\author[0000-0003-3939-3297]{Robyn E. Sanderson}
\affil{\upenn}
\affil{\cca}

\author[0000-0002-8354-7356]{Arpit Arora}
\affil{\upenn}

\author[0000-0002-6993-0826]{Emily C. Cunningham}
\altaffiliation{NASA Hubble Fellow}
\affiliation{\cca}
\affiliation{\columbia}

\author[0000-0002-9306-1704]{Jay Baptista}
\affil{\yale}
\affil{\ifa}

\begin{abstract}
Recent studies have suggested that the Milky Way (MW)'s Dark Matter (DM) halo may be significantly tilted with respect to its central stellar disk, a feature that might be linked to its formation history. In this work, we demonstrate a method of constraining the orientation of the minor axis of the DM halo using the angle and frequency variables. This method is complementary to other traditional techniques, such as orbit fitting. We first test the method using a simulated tidal stream evolving in a realistic environment inside an MW-mass host from the FIRE cosmological simulation, showing that the theoretical description of a stream in the action-angle-frequency formalism still holds for a realistic dwarf galaxy stream in a cosmological potential. Utilizing the slopes of the line in angle and frequency space, we show that the correct rotation frame yields a minimal slope difference, allowing us to put a constraint on the minor axis location. Finally, we apply this method to the Sagittarius stream's leading arm. We report that the MW's DM halo is oblate with the flattening parameter in the potential $q\sim0.7-0.9$ and the minor axis pointing toward $(\ell,b) = (42\degree,48\degree)$. Our constraint on the minor axis location is weak and disagrees with the estimates from other works; we argue that the inconsistency can be attributed in part to the observational uncertainties and in part to the influence of the Large Magellanic Cloud.  

\end{abstract}

\section{Introduction}

In the standard $\Lambda$CDM universe, structures are formed in a hierarchical fashion. Less massive structures such as dwarf galaxies and globular clusters can be consumed by a host galaxy such as the Milky Way (MW). As they fall into the deeper potential well, their stars are tidally stripped and stretched along the orbit, creating tidal streams. Indeed, numerous detectable tidal streams have been discovered within the MW \citep[e.g.][]{Shipp2018, galstream, li2022}. Of these, one of the earliest to be discovered was the Sagittarius (Sgr) stream. With SDSS and 2MASS, \cite{Newberg2002} and \cite{Majewski2003} identified the tidal tails associated with the Sagittarius dwarf galaxy \cite{ibata1994} as it is being consumed by the MW. Two tidal tails were detected, the leading arm in the northern Galactic hemisphere and the trailing arm in the south.

Tidal streams serve as a powerful tool to constrain the properties of their host galaxies. In the past, the gold standard is to analyze the streams in position and velocity space. Since the discovery of the Sgr stream, numerous studies have tried to constrain the properties of the MW's DM and stellar halo, as well as its underlying DM distribution, by fitting the modeling positions, line-of-sight velocities, and distances to the observed properties of the stream stars. Many tensions have arisen. The prolate halo shape is preferred to describe the velocity structure of the leading arm \citep{helmi2004, law2005}, while the Sgr spatial appearance is better explained by an oblate halo \citep{johnston2005}. Soon later, \cite{law2010} explores a more general class of arbitrarily oriented triaxial potentials. The best-fit model prefers a significantly rotated halo with the minor axis pointing along the Galactic plane. Such configuration has been argued to be unlikely due to its dynamical instability in nature \citep{debattista2013}. Thanks to \textit{Gaia}\citep{gaia-mission} and other surveys, we can now access complete 6-D data of tidal streams that open up new analysis methods. Because of the similarity in their stellar orbits, an action-angle framework ($J_i$, $\theta_i$, $\Omega_i$) is frequently used to analyze tidal streams as they are often simplified in this choice of coordinates. In this framework, stars in a tidal stream (under certain assumptions about the galactic potential and its evolution) are clustered in the action space, $J_i$. This allows one to constrain the best-fit galactic potential that translates to the most clustered actions, using either individual streams or multiple streams simultaneously \citep{sanderson2017, reino2021}. Beyond actions, one can also exploit the angle ($\theta_i$) and frequency ($\Omega_i$) variables to constrain the galactic potential. \cite{sanders2013b} shows that a given stream appears as line overdensities in both the angle and frequency space, with the best-fit galactic potential producing matching slopes in both of these spaces. They also test this method on a mock tidal stream simulated under a two-parameter logarithmic galactic potential and show that they can recover the true parameters for a stream with $\sim500$ stars.

However, the Galactic potential is a complex, time-evolving system as it directly traces the underlying mass distribution of the MW in real time. Specifically, the DM halo is twisted and tilted with respect to the central stellar disk and assumes a non-trivial shape. This has been shown through various means such as stream-orbit fitting \citep[e.g.][]{law2010, erkal2019, vasiliev2021} and inferring from the plane of satellites \citep{shao2020}. The studies of the stellar halo have also suggested a large-scale tilt at $r_{gal}\sim 30$ kpc, hinting that the DM halo may possess the same nature \citep[e.g.,][]{han2022, han2022b}. The reason for the tilted halo can be explained by the interplay between the long-lived structures from past mergers such as the \textit{Gaia} Sausage Enceladus \citep{Helmi2018,Belokurov2018} and the transient and collective responses induced by the Large Magellanic Cloud (LMC) in both the stellar and DM halos \citep{garavito2019,cunningham2020,garavito2021,conroy2021}. Moreover, these constraints are not very consistent as most of these methods can only probe the DM halo within certain radial ranges. To get a complete picture of the DM halo's structure, complementary methods of measuring its tilt are needed.

Cosmological hydrodynamical simulations can also be used to study the galactic halo and tidal debris around MW-mass hosts. One important finding reveals that a tilted halo is not uncommon. For example, \cite{emami2021} shows that only $32\%$ of the MW-mass galaxies in the Illustris TNG50 simulation have a \emph{simple} halo (the halo orientation is almost fixed). The remaining either have a \emph{twisted} (gradual rotation as a function of radius) or \emph{stretched} (swapping of principal axes at certain radii) halo. Zoomed cosmological-hydrodynamical simulations in the Latte suite \citep{Wetzel2016} also display diverging halo and stellar disk axes beyond 30 kpc (Baptista et al. in prep.).These zoomed simulations also have enough particle resolution for studying dwarf galaxy streams down to $\sim 10^6$\Msun in stellar mass \citep{Panithanpaisal2021}.

In this manuscript, we re-evaluate the constraining power of the angle and frequency method first presented in \cite{sanders2013b} in the context of realistic cosmological potentials. Our goal is to show that, even with the complex nature of the galactic potential, this method can still give valuable information to constrain the DM halo tilt to some extent. However, unlike \cite{sanders2013b}, we do not try to determine the actual shape (constraining the parameters) of the galactic potential since it cannot be described effectively in analytical form. Instead, we aim to constrain its orientation using an axisymmetric approximation to the potential. The rest of the manuscript is organized as follows. In \S\ref{sec:method} we describe the action-angle framework and the method in detail. In \S\ref{sec:sim_data} we test the method on a simulated stream from cosmological simulations of galaxy formation. In \S\ref{sec:sag_data} we apply the method to the Sgr stream data. We discuss our results in \S\ref{sec:discussion}. Finally, we summarize the content of this manuscript in \S\ref{sec:summary}.



\section{Method}\label{sec:method}
\subsection{action-angle framework}\label{sec:action-angle}
The action-angle formalism provides the simplest description of a stream \citep{10.1046/j.1365-8711.1999.02616.x, 10.1046/j.1365-8711.1999.02690.x}. Given 6D phase-space information and a model of the global potential under which Hamilton's equations are separable, we may find a canonical transformation to action-angle coordinates, $(J_i, \theta_i)$. The actions $J_i$ and angles $\theta_i$ obey
\begin{equation}
    J_i = \textrm{const.}
\end{equation}
\begin{equation}
    \theta_i(t) = \theta_i(0) + \Omega_i t,
\end{equation}
where $\Omega_i = \partial\mathcal{H}/\partial J_i$ are the associated frequencies of a star with Hamiltonian $\mathcal{H}$ and $t$ is the time since the star was tidally stripped from the progenitor. In this formulation, we assume that stream stars are evolving independently under an external static or adiabatically time-evolving galactic potential and that the gravitational influence of the progenitor can be neglected once each stream star has been stripped. In other words, the actions of the stream stars are fixed at the time they were stripped. Since the $J_i$ are constant, a star on a bound orbit moves only on a three-dimensional torus in the $\theta_i$ subspace, which is $2\pi$-periodic.

Streams form because stripped stars are on slightly different orbits than their progenitor \citep[e.g.,][]{sanders2013a}. In the action-angle formalism, the differences in angles between a single stream star and the progenitor follow
\begin{equation}
    \Delta \theta_i \approx \Delta \Omega_i t,
\end{equation}
where $\Delta \Omega_i$ are the differences in frequencies between the star and the progenitor. If the frequency differences are small, we can Taylor expand the frequencies of the star around the frequencies of the progenitor $\Omega_{i,0}$, and hence,
\begin{equation}\label{eq:stream_growth}
    \Delta \theta_i \approx \Delta \Omega_i t \approx (\mathcal{D}_{ij}\Delta J_j) t,
\end{equation}
where \textbf{$\mathcal{D}_{ij}$} is the Hessian matrix is given by
\begin{equation}
    \mathcal{D}_{ij} = \frac{\partial^2 \mathcal{H}}{\partial J_i \partial J_j}.
\end{equation}
Since the differences in actions are fixed at the tidal stripping time, the Hessian exclusively determines the directions the stream spreads. The Hessian has three principal eigenvectors, $\lambda_1$, $\lambda_2$ and $\lambda_3$. For a stream, one of the eigenvalues is much larger than the others (e.g. $\lambda_1 \gg \lambda_2, \lambda_3$). This causes the stream to grow in the direction parallel to the corresponding principal eigenvector, $\hat{e}_1$, in both the angle and frequency spaces following Equation \ref{eq:stream_growth} \citep{sanders2013a}.

\subsection{the algorithm for constraining the Galactic potential}\label{sec:algo}
To compute action, angle, and frequency variables, we need the kinematic information of each star as well as a model for the galactic potential. Following the argument in \cite{sanders2013b}, in the correct potential, the directions of the spreads in both the angle and frequency spaces should best align. Hence, invoking different trial potentials and measuring the degree of misalignment allows us to constrain the true galactic potential. However, as opposed to \cite{sanders2013b}, we do not focus on finding the best-fit parameters for the galactic potential (the halo shape). Instead, we only try to constrain the symmetry axis of the galactic halo because of a few reasons. First, modeling the accurate galactic potential is non-trivial since it is time-evolving, radially-dependent, and non-axisymmetric. Even if we could model the galactic potential perfectly, it would still introduce biases in our action computations (even with error-free kinematic information) since they can only be approximated. Moreover, the time evolution may or may not be adiabatic at all radii. Second, the MW contains numerous subhalos which can interact with tidal streams. Stream-subhalo interactions can further impact the action computations. Their effects are not fully quantified and are beyond the scope of this manuscript. Specifically, we only consider an axisymmetric potential model since the tilt can be unambiguously described just by specifying the minor axis location. \cite{sanders2013b} only test the method on a simulated GC-like stream evolving under a static, smooth, two-parameter logarithmic potential, and do not generalize to a more realistic setup. Our goal is to show that even with the non-realistic assumption of the Galactic potential (the smooth axisymmetric model), applying this method to a realistic simulated tidal stream can still constrain the axes of symmetry of the MW halo, hence quantifying its tilt. We will apply the method to a realistic simulated stream from a cosmological hydrodynamical simulation of galaxy formation and then to the Sgr stream. We will show that the slope difference of the spread in angle and frequency spaces will still be minimized under the rotating frame that best aligns with the true halo symmetry axes. 

In practice, many python packages are capable of estimating action, angle, and frequency variables. Specifically, we perform action computations using \texttt{AGAMA} \citep{2019MNRAS.482.1525V}. \texttt{AGAMA} provides various ways of converting between positions and velocities to actions and angles. However, for the choices of trial potentials that we will be using (axisymmetric), \texttt{AGAMA} utilizes the Stack\"el fudge approximation \citep{2012MNRAS.426.1324B}. The errors associated with this approximation is largest for box orbits and are much smaller for loop orbits which describe most of the stream stars of the simulated stream (\S\ref{sec:sim_data}) and Sgr stream (\S\ref{sec:sag_data}). The Stack\"el fudge method assumes that the motion is integrable and can be described locally by a Stack\"el potential. The Stack\"el potential is a special class of potentials that are separable ellipsoidal coordinates, allowing for easy and more accurate action computations. For an axisymmetric potential model, the most convenient choice of actions and angles are the triplet $\{J_r, J_z, J_\phi\}$ and $\{\theta_r, \theta_z, \theta_\phi\}$. The first two actions $J_r$ and $J_z$ describe the amplitudes of the oscillations in the cylindrical radial and vertical directions. The third action $J_\phi$ is the conserved z-component of the orbital angular momentum $L_z$. The three angles describe the corresponding 2$\pi$-periodic phases of these oscillations.

Since the spread in angle and frequency spaces lies along a line (one eigenvalue of the Hessian is much larger than the other two), its slope can be described by a 3-D vector pointing parallel to this line. However, we choose to describe this line by its slopes in three 2-D projected spaces due to the computation time of the automatic line detection algorithm introduced in \S\ref{sec:hough}. This will also allow for uncertainty estimation described in \S\ref{sec:m12i_uncertainties}. We define the 2-D projected slopes in angle space as $\{n_{rz}, n_{r\phi}, n_{z\phi}\}$, where they represent the slope in $r-z$, $r-\phi$ and $z-\phi$ projection, respectively. Similarly, the slopes in frequency space are given by $\{m_{rz}, m_{r\phi}, m_{z\phi}\}$. Under the correct rotating frame, we expect all the slopes to align (i.e. $n_{ij}/m_{ij} \sim 1$ for $i,j \in \{r,z,\phi\}$). We thus define the \emph{slope difference} $\Delta$ to be the L1 norm of the projected misalignments:
\begin{equation}\label{eq:slope_diff}
    \Delta = \left| 1 - \frac{n_{rz}}{m_{rz}}\right| + \left| 1 - \frac{n_{r\phi}}{m_{r\phi}}\right| + \left| 1 - \frac{n_{z\phi}}{m_{z\phi}}\right|,
\end{equation}
where the best-fit orientation will minimize $\Delta$. Given a stream dataset containing the positions and velocities of its member stars, we perform the following steps to find the best-fit orientation,
\begin{enumerate}
    \itemsep0em 
    \item pick the best axisymmetric Galactic potential $\Phi_g$ that describes the radial profile of the host, assuming alignment with the disk plane;
    \item pick a trial Galactic longitude and latitude $(\ell_{trial},b_{trial})$, then rotate the coordinate frame such that the positive z-axis of the potential now points in this direction;
    \item compute the angle and frequency variables in the rotated frame assuming the Galactic potential $\Phi_g$;
    \item estimate all the slopes $\{n_{rz}, n_{r\phi}, n_{z\phi}\}$ and $\{m_{rz}, m_{r\phi}, m_{z\phi}\}$ using the line detection algorithm discussed below in \S\ref{sec:hough}, then compute $\Delta$;
    \item repeat step 2 until the entire hemisphere of possible $(\ell_{trial},b_{trial})$ is densely sampled.
\end{enumerate}


\subsection{line detection} \label{sec:hough}
\begin{figure}[t]
\plotone{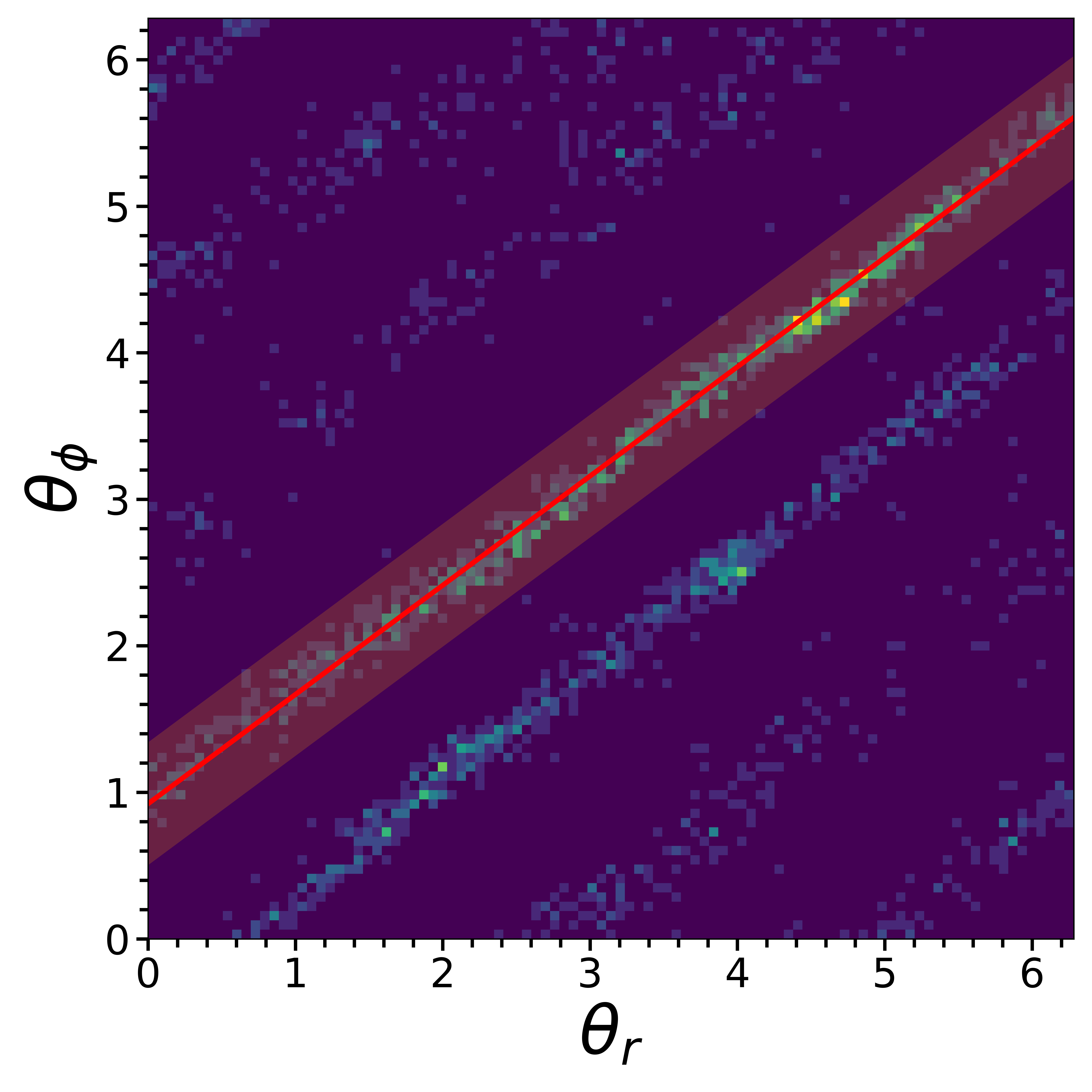}
\caption{Example of a line detection on a simulated stream in a 2-D angle space projection $\theta_r-\theta_\phi$. This long stream contains multiple wraps in this projection. The most prominent line detected and its associated width is shown by the orange band. We then perform a linear least-square fit on stream stars that are inside this band. The best-fit line is shown in red.  \label{fig:line_detection}}
\end{figure}

To automate the process of estimating the slopes, we use a line detection algorithm. A simple, yet effective line detection algorithm is Hough transform \citep{10.1145/361237.361242}. This algorithm can be modified \citep[e.g.][]{2020ascl.soft03005C} and can be applied directly to photometric data to detect real streams \citep[e.g.][]{2022ApJ...926..166P}; in this work we build our own implementation of Hough transform and apply it to detect lines in angle and frequency space. The equation of a straight line in 2-D Cartesian coordinates is given by,
\begin{equation}
    s = x\cos\phi + y\sin\phi,
\end{equation}
where $s$ is the distance between the origin and the closest point along the line and $\phi$ is the angle between the x-axis and the line connecting the origin to that closest point. In this framework, for a given data point $(x,y)$, $s$ is computed from a set of $\phi$'s with values ranging from 0 to $\pi$ in uniform spacings. The most prominent line then corresponds to the peak in the $(s,\phi)$ distribution.

For a given 2-D angle or frequency projection, we loop through each star one by one, computing the $(s,\phi)$ value pairs for $\phi$ ranging from 0 to 180 degrees in uniform steps of $1\degree$. This choice of step size translates to $\sim 1\degree$ sensitivity in the detected slope and is chosen so that the subsequent computations remain manageable. Next, for each $\phi$, we use a 1D Epanechnikov kernel density estimator with bandwidth 0.5 degrees\footnote{\code{scikit-learn} (\code{kernel=Epanechikov}; \code{bandwidth=0.5})} to determine the smooth density distribution and record the peak density value.  The highest peak in density $(s_{peak},\phi_{peak})$ determines the most prominent line in the 2-d angle or frequency projection. With the detected line, we then perform a linear regression on the stars that are within its width to accurately measure its slope. Specifically, the width of the line is determined relative to the peak at 0.75 depth (i.e. slightly larger than the FWHM). The slope of this best-fit line in each projection is then used to compute $\Delta$. Figure \ref{fig:line_detection} shows an example of the line detected by this algorithm. The example stream is long and contains multiple wraps in the $\theta_r-\theta_\phi$ projection. We successfully detect the most prominent line and its associated width which is shown as the orange band. We then perform the linear least-square fit on the stream stars that are inside this band, obtaining the best-fit line shown in red.

\section{testing on simulated stream}\label{sec:sim_data}
In this section, we test the method by applying it to a simulated stream that forms and evolves in the full cosmological hydrodynamical setting. 

\begin{figure*}
\plotone{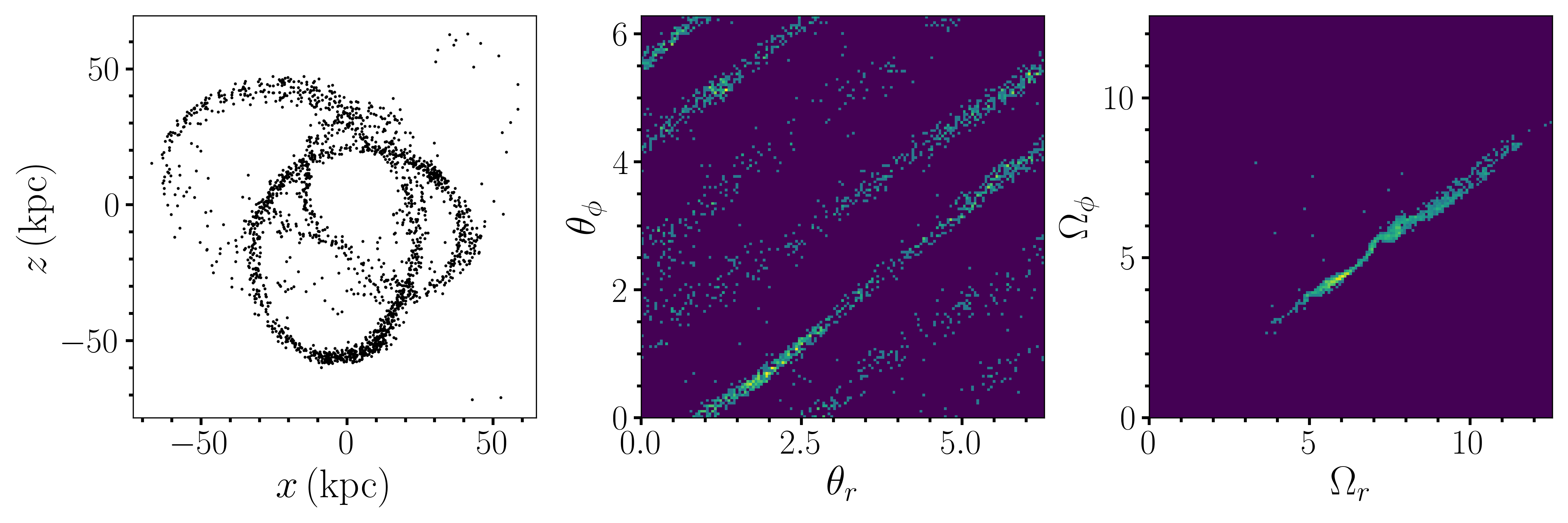}
\caption{Different 2D projections of the simulated stream from FIRE cosmological simulations that we use to test our method. This stream has $\sim 2000$ star particles with stellar mass of $1.0\times10^7\Msun$. The projections are real space $x-z$ projection (left), angle space $\theta_r-\theta_\phi$ projection (middle) and frequency space $\Omega_r-\Omega_\phi$ projection (right). Everything is shown/computed in the default host principal axis frame at the present day, where the $z$ direction points perpendicular to the disk. The potential used in angle and frequency computations is the multipole potential described in \S\ref{sec:pot1}. \label{fig:sim_stream}}
\end{figure*}

\subsection{simulation data}
We select our test stream from the suite of zoomed, cosmological-hydrodynamical simulations of MW-mass galaxies in the Feedback In Realistic Environments (FIRE) project\footnote{See the project website at \href{http://fire.northwestern.edu}{http://fire.northwestern.edu}}. These simulations are run with \code{Gizmo} \citep{Hopkins2015} with star formation and stellar feedback consistent with stellar evolution models like \texttt{STARBURST99} \citep{Leitherer1999}. This produces MW-mass hosts with properties that are broadly consistent with properties of the MW \citep{Ma2017, 2018MNRAS.480..800H, 2019MNRAS.485.5073D, 2020ApJS..246....6S}. The MW-mass hosts are included in the Latte suite \citep{Wetzel2016} and ELVIS suite \citep{Garrison-Kimmel2019} with a total of 13 hosts combined. All the halos were simulated with $\Lambda$CDM cosmology at particle mass resolution of 3500--7100 $\Msun$ and spatial resolution of 1--4 pc for star and gas particles. The DM particles are simulated at slightly lower resolution at 18,000--35,000 $\Msun$ per DM particle and 40 pc softening. This resolution allows us to study tidal streams around the host galaxy with stellar mass down to that of MW's classical dSphs ($\sim 10^6 \Msun$).

\cite{Panithanpaisal2021} identified $\sim 100$ streams with stellar mass between $10^{5.5}\Msun$ and $10^9\Msun$ across these 13 hosts and publicly released the data \citep{2022arXiv220206969W}. The streams were selected by first tracking luminous substructures between 2.7 and 6.5 Gyr ago that are within the virial radius of the host halo at the present day. We then apply several criteria (on the number of star particles, pairwise distances between star particles, and velocity dispersion) to eliminate substructures that are still not disrupted (dwarf galaxies) or structures that are already phase-mixed at the present day. Refer to Section 3 of \cite{Panithanpaisal2021} for more detail.

Most of the streams have relatively low stellar masses containing fewer than 500 star particles. The remaining streams also have a variety of orbits ranging from very radial to almost circular, spanning a wide range of distances. Some still contain a bound progenitor that survives to the present day. Within this wide variety, we select a test stream that we consider to be a suitable Sgr stream analog. First, like Sgr, the test stream is in a very polar orbit with the orbital plane being almost perpendicular to the disk. Second, the total stellar mass of this stream is thought to be comparable to the stellar mass of the Sgr stream. The total mass of the Sgr remnant is estimated to be $\sim 10^8\Msun$ \citep[e.g.][]{gibbons2017,vasiliev2020} which is around half its mass at infall \citep{2010ApJ...712..516N} (the mass estimated from the metallicity-mass relationship can be slightly higher). Finally, the stream is mostly orbiting at $\sim50$ kpc from the host which is roughly the same distance as the portion of the Sgr stream with available 6D data. However, unlike the Sgr stream, the test stream does not have a surviving progenitor as it was completely disrupted $\sim 2$Gyr ago (redshift $z=0.15$), which eliminates the need to mask out the progenitor during action computations. Figure \ref{fig:sim_stream} (left) shows a scatter plot of the test stream in real space $x-z$ projection with respect to the default host principal axis frame, where the $z$ direction points perpendicular to the stellar disk. This stream contains $\sim 2000$ star particles with total stellar mass of $\sim 2.0\times 10^7 \Msun$.

\begin{figure*}[t]
\plotone{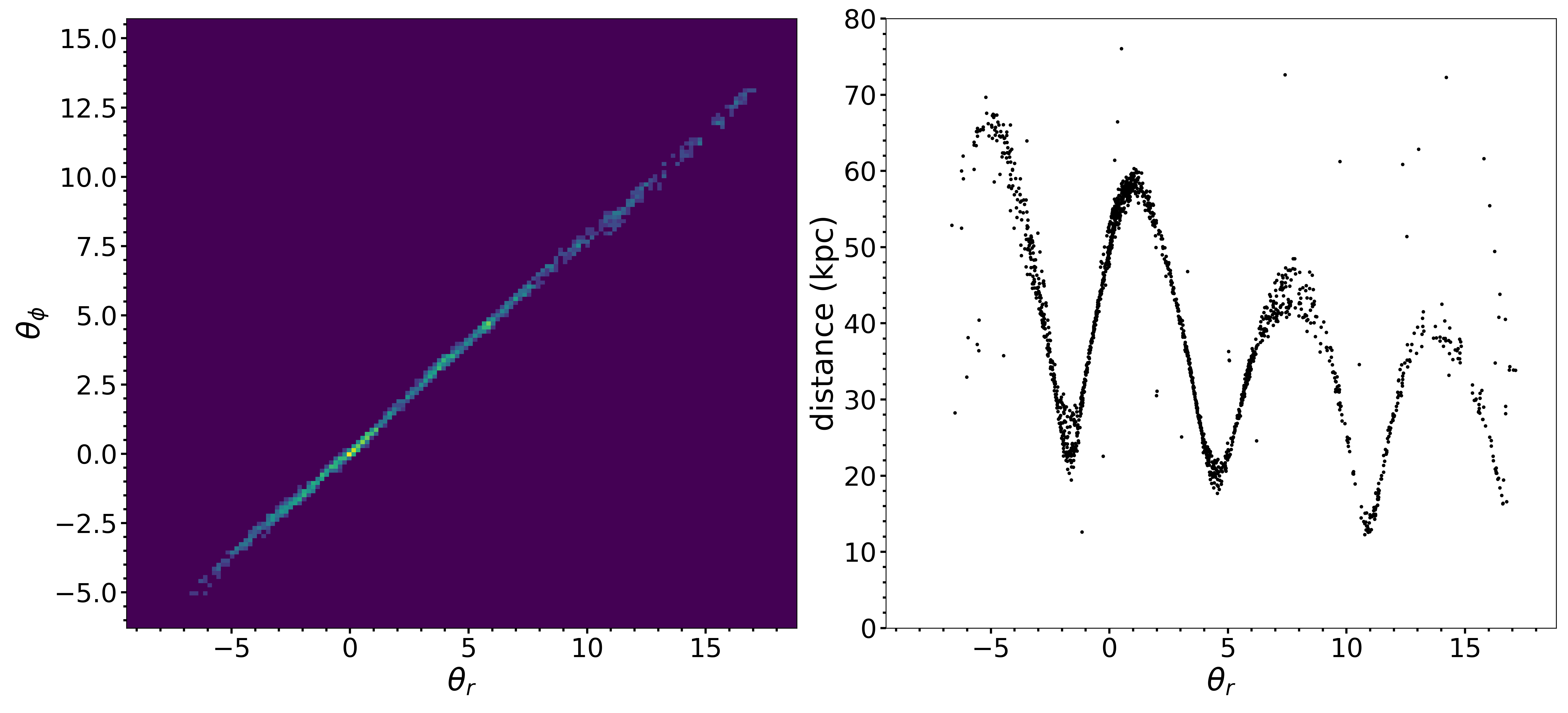}
\caption{Left: A 2-D histogram of the unwrapped stream in $\theta_r - \theta_\phi$ projection, compared to the wrapped stream in Figure \ref{fig:sim_stream} (middle panel). Right: central distance of each stream star as a function of the unwrapped $\theta_r$. This shows how well we can unwrap the stream in this particular case, although not all cases are as clean as this one. The unwrapped angle increases monotonically along the stream. \label{fig:unwrap}}
\end{figure*}

\subsection{potential model}\label{sec:pot1}
For the simulated stream, we use an axisymmetric potential fit using basis function expansions. Specifically, we use the multipole potential model of the host \texttt{m12i} estimated at a present-day snapshot from \cite{arora2022}. In this model, the DM halo and hot gas ($T_{gas}>10^{4.5}$K) are described by the linear combination of spherical harmonic basis $Y_\ell^m(\theta,\phi)$, while the stellar and cold gas ($T_{gas}<10^{4.5}$K) are described by a set of Fourier harmonics in azimuthal coordinates. For axisymmetric models, the only non-vanishing modes are the zeroth pole for azimuthal expansion and even $\ell$ pole with $m=0$. We only utilize the lower order $\ell \leq 4$, as this has been shown to be sufficient to preserve the coherence of streams in action space if we consider time-evolution \citep{arora2022}. Assuming this potential, we then use \texttt{AGAMA} to compute the angle and frequency variables for each stream star. Figure \ref{fig:sim_stream} shows 2-D histograms of the simulated stream in angle space $\theta_r-\theta_\phi$ projection (middle) and frequency space $\Omega_r-\Omega_\phi$ projection (right), also in the default principal axes frame. The colored bins contain a non-zero number of stream stars, while the dark background contains no stars.

\subsection{the degeneracy issue in determining $\{n_{rz}, n_{r\phi}, n_{z\phi}\}$}\label{sec:degeneracies}
The stream in angle and frequency space spreads into line overdensities as we described in \S\ref{sec:action-angle}. Due to the perfect knowledge of the stream members, the stream contains multiple wraps, spanning all possible orbital phases. With the periodic boundary condition of the angle space, the stream thus appears as multiple line overdensities wrapping around in any 2-D projection. This complicates our effort to determine the slopes $\{n_{rz}, n_{r\phi}, n_{z\phi}\}$ because of the degeneracies in slope between wraps. Ideally, these line overdensities have the same slope, pointing along the first principal eigenvector $\hat{e}_1$ of the Hessian matrix. However, in practice these lines will have slightly different slopes due to multiple reasons: the potential approximation, the use of the Stack\"el fudge approximation for action computations, and the assumption that the potential is adiabatically evolving. There are also physical mechanisms such as subhalo interactions that can slightly nudge a star out of its place.

We first considered addressing the degeneracy issue by \emph{unwrapping} the stream in angle space. To summarize, after finding the most prominent line (corresponding to ($s_{peak},\phi_{peak}$)) with the Hough transform, we can subsequently search for the local maxima at fixed $\phi_{peak}$. This is equivalent to searching for all possible lines at a fixed slope. A threshold value just above the noise level is required such that peaks that are below the threshold are discarded. All the ($s_{peak}^{local},\phi_{peak}$) are then considered detections. Next, we connect all the lines by exploiting the periodic boundary condition in the angle projection (i.e. if a line terminates at $(\theta_r, \theta_\phi) = (2\pi, \theta_\phi^0)$, it will reappear on the other side at $(\theta_r, \theta_\phi) = (0, \theta_\phi^0)$). Starting from the most prominent line as the stem, we progressively connected more lines on both sides, shifting $(\theta_r, \theta_\phi)$ of the members of the newly attached lines with appropriate offsets, until all the lines are used. Figure \ref{fig:unwrap} shows an example of the unwrapped stream. The left panel shows the 2-D histogram in $\theta_r - \theta_\phi$ projection, compared to the wrapped stream in Figure \ref{fig:sim_stream} (middle panel). The right panel shows the distance of the stream star with respect to the unwrapped $\theta_r$, which highlights how well we can unwrap the stream in this particular case by using just the angle space alone with no extra information such as total energy or distance of each stream star and confirms that the stream in cosmological settings still follows our theoretical expectation in action, angle and frequency space.

Unfortunately, stream unwrapping also introduces another layer of uncertainties into our slope measurements since there are a lot of corner cases that prevent full automation. For example, some diffuse lines can evade our detection, causing problems with unwrapping by disconnecting the neighboring wraps. Moreover, the estimated frequencies can be near resonances such that some line overdensities in angle space overlap. These require manual intervention to fix on a case-by-case basis, which is not feasible since we will later explore the angle and frequency variables for over 1000 different possible rotations of the coordinate frame. Hence, we instead determine $\{n_{rz}, n_{r\phi}, n_{z\phi}\}$ as the slope of the line that contains the most stars. The choice is also physically motivated since this should correspond to the part of the stream that has been stripped most recently and is least likely to have experienced subhalo interactions or potential evolution since stripping.

\begin{figure*}[t]
\plotone{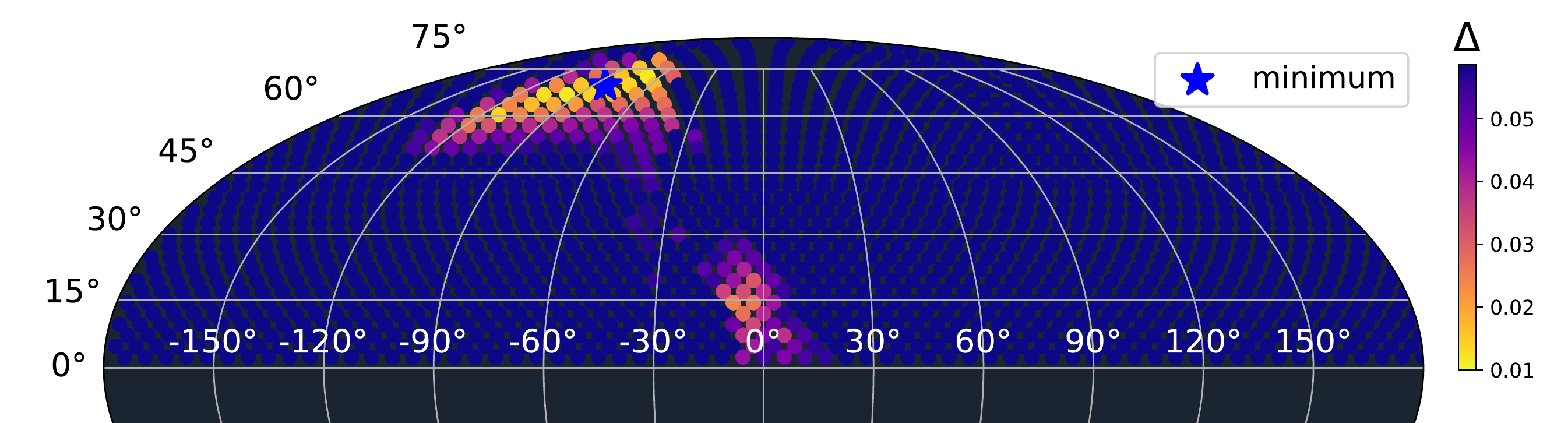}
\caption{Results for the simulated stream, following the procedures outlined in \S\ref{sec:algo}. We compute the slope difference $\Delta$ for each trial Galactic latitude and longitude location $(b_{trial},\ell_{trial})$. The angle and frequency variables are computed by assuming a fixed axisymmetric multipole potential estimated directly from the simulation at the present-day snapshot. In this Figure, all the trial rotations are shown with filled circular markers, with the color representing estimated $\Delta$. The blue star marks the location that minimizes $\Delta$ at $(\ell,b)=(-83.6\degree, 63.4\degree)$. The true DM halo minor axis at 50 kpc (estimated by constructing the moment of inertia tensor using DM particles in the simulations) is at $(\ell_{true},b_{true})=(-96.5\degree,65.5\degree)$.\label{fig:sim_slope_diff}}
\end{figure*}

\subsection{results}\label{sec:results_sim}
We follow the procedures outlined in \S\ref{sec:algo} to find the region of $(\ell, b)$ rotation space that produces the minimal slope difference $\Delta$. First, the chosen galactic potential is described in \S\ref{sec:pot1}. Second, we split the surface of a sphere into pixels of equal areas using \code{healpy} \citep{2005ApJ...622..759G, Zonca2019}. We set \code{nside=16}, so that each pixel is spanning $\sim 3.6\degree$ on each side. The centers of all the pixels are chosen to be the trial rotations ($\ell_{trial},b_{trial}$). Since our potential is axisymmetric, we also restrict to $b_{trial} > 0$. This results in a total of 1504 trial rotations for our initial grid. For each trial rotation, we construct a rotation matrix that is then applied to the position and velocity vectors of the stream stars and finally estimate the slope difference $\Delta$.

Figure \ref{fig:sim_slope_diff} shows the map of $\Delta$ as a function of ($\ell_{trial},b_{trial}$) for the simulated stream. The filled circular markers show all the trial rotations in the galactic latitude and galactic longitude coordinate system. The estimated slope difference $\Delta$ associated with each trial rotation is shown by marker color, with the minimal $\Delta$ regions having lighter colors. The trial rotation with minimum $\Delta$ is marked by the blue star. At first glance, there are two distinct regions that yield minimal $\Delta$. \textbf{Region 1} is centered around the  global minimum ($\Delta = 0.010$)at $(\ell,b)=(-83.6\degree, 63.4\degree)$, while \textbf{Region 2} is near the Galactic equator at $(\ell,b)\approx(-5\degree, 15\degree)$ with $\Delta\sim0.025$. To attain such minimal $\Delta$, the slopes of the line overdensity in angle and frequency space have to match up to within 1 percent in all projections ($r-z, r-\phi, z-\phi$).

To further understand these results, we compare them to the true DM halo minor axis location. We can estimate the true shape and orientation of the DM halo by directly selecting the DM particles within a certain radius and estimating the reduced moment of inertia tensor \citep[e.g.,][]{allgood2006}, which is given by
\begin{equation}\label{eq:moi}
    I_{ij} = \sum_n \frac{x_{i,n}x_{j,n}}{r_n^2},
\end{equation}
where $i,j=1,2,3$, $x_{i,n}$ is the position of the $n^{th}$ star particle along the $i^{th}$ direction, and $r_n$ is the ellipsoidal distance. The most general halo shape is a triaxial, described by three parameters $(a,b,c)$ that represents the half-length of the major, intermediate and minor axis, respectively. The direction and length of these parameters are related to the eigenvalues and eigenvectors of $I_{ij}$. We select DM particles within 50 kpc (similar to the distances of stream stars) and construct $I_{ij}$ following Equation \ref{eq:moi}. Diagonalizing the tensor, we find $b/c =0.949$ and $c/a=0.863$. This is closer to an oblate than a prolate shape. The direction of the minor axis (the eigenvector of lowest eigenvalue) points towards $(\ell_{true},b_{true})=(-96.5\degree,65.5\degree)$, consistent with \textbf{Region 1}.

\subsection{uncertainty estimation}\label{sec:m12i_uncertainties}
Our slope measurements are not exact, due in part to our chosen strategy for dealing with the angle space degeneracies (\S\ref{sec:degeneracies}) and in part to our method for determining the linewidth before performing the least-square fit (\S\ref{sec:hough}). In fact, it is counter-productive to try to measure the most accurate slopes since our angle and frequency variables are not entirely faithful due to the potential choice (see \S\ref{sec:pot1}) and the Stack\"el fudge approximation used by \code{AGAMA}. The more important thing is to have the means to quantify the uncertainties in our slope measurements and, hence, the uncertainty in $\Delta$, that encompasses all these numerical and systematic effects.

For a straight line in 3-D space, we can describe its full 3-D slope (its unit vector) by measuring the slopes in just 2 different 2-D projections. In other words, we only need to measure the slopes in two different 2-D projections in order to infer the slope in the third projection, 
\begin{equation}
    n_{z\phi} = n_{r\phi}/n_{rz},
\end{equation}
and similarly,
\begin{equation}
    m_{z\phi} = m_{r\phi}/m_{rz}.
\end{equation}
We intentionally define $\Delta$ as described in Equation \ref{eq:slope_diff} even though not all the terms are independent to impose a self-consistency check of our implementation of the Hough transform and slope measurement. In some cases, one of the terms in Equation \ref{eq:slope_diff} blows up, signaling a problem such as when the frequencies are almost in resonance, causing multiple lines to be detected as one by our line detection algorithm. We quantify the uncertainties of the slope measurements in angle and frequency space in terms of the consistency between the slopes measured in the three projections:
\begin{equation}
    \sigma_n = \left| \frac{n_{r\phi}}{n_{rz}} - n_{z\phi}\right|, 
\end{equation}
and 
\begin{equation}
    \sigma_m = \left| \frac{m_{r\phi}}{m_{rz}} - m_{z\phi}\right|, 
\end{equation}
respectively. This definition encompasses both numerical uncertainty from our computational methods and systematic uncertainty related to the limitations of our potential models. Then, the uncertainty in $\Delta$ is obtained by quadrature:
\begin{equation}\label{eq:sigma_delta}
    \sigma_\Delta^2 = \sigma_{rz}^2 + \sigma_{r\phi}^2 + \sigma_{z\phi}^2, 
\end{equation}
where $\sigma_{r\phi}$ is defined as
\begin{equation}
    \sigma_{r\phi}^2 = \left( \frac{\partial\Delta}{\partial n_{r\phi}} \right)^2\sigma_n^2 + \left( \frac{\partial\Delta}{\partial m_{r\phi}} \right)^2\sigma_m^2,
\end{equation}
and similarly for $\sigma_{rz}$ and $\sigma_{z\phi}$.  We estimate $\sigma_\Delta$ for the trial rotations within \textbf{Region 1} to be $\sim 0.011$ and for \textbf{Region 2} to be $\sim 0.060$, hence we reject \textbf{Region 2}.

\begin{figure}[t]
\plotone{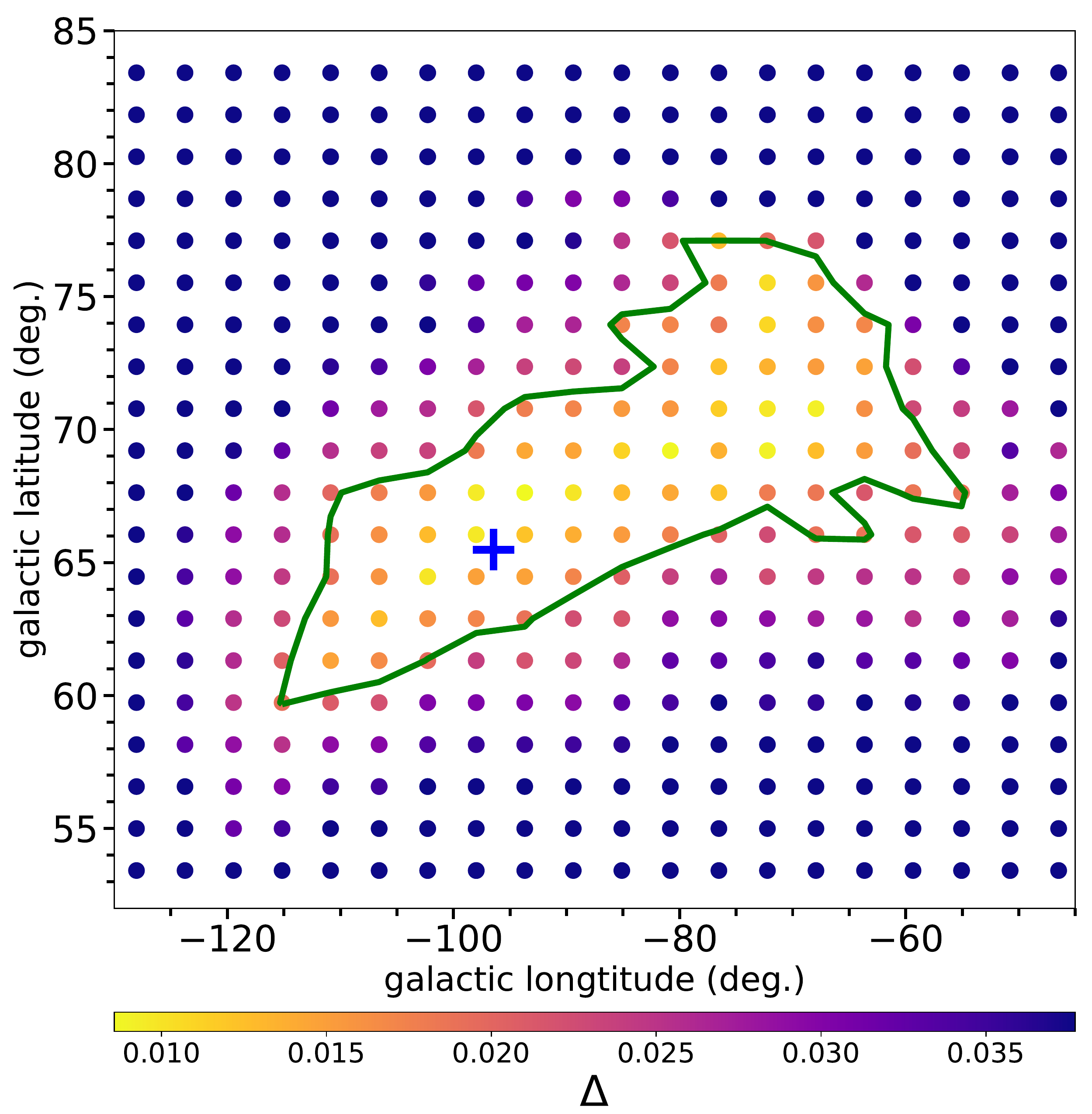}
\caption{$\Delta$ estimates for the simulated stream, calculated on a zoom-in rectangular grid around the minimum region shown in Figure \ref{fig:sim_slope_diff}. The scatter points are individual trial rotations with the color showing the estimated slope difference $\Delta$. The $1\sigma$ uncertainty contour around the minimal region is shown in green. The blue cross shows the true minor axis location of the DM halo at 50 kpc. \label{fig:m12i_uncertainty}}
\end{figure}

We set up a refined rectangular grid around $\textbf{Region 1}$ and compute $\Delta$ and $\sigma_\Delta$ for each point in this more finely sampled region of parameter space, following Equations \ref{eq:slope_diff} and \ref{eq:sigma_delta} respectively. Reassuringly, $\sigma_\Delta$ is smooth and monotonically increasing within our zoom-in region. We define the contour encompassing the minimum and its corresponding uncertainty to be 
\begin{equation}\label{eq:uncertainty_countour}
    \Delta < \Delta_{min} +\sigma_{\Delta_{min}},
\end{equation}
where $\Delta_{min}$ is the minimum $\Delta$ within our zoom-in region and $\sigma_{\Delta_{min}}$ is its associated uncertainty. We find $\Delta_{min} = 0.008$ and $\sigma_{\Delta_{min}} = 0.011$, hence our $1\sigma$ uncertainty contour corresponds to $\Delta < 0.019$. The zoom-in results are shown in Figure \ref{fig:m12i_uncertainty}. The scatter points are the individual trial rotations, with the colors showing their associated $\sigma_\Delta$. The $1\sigma$ uncertainty contour is drawn in green. The true minor axis location of the DM halo at 50 kpc is marked by the blue cross, which is within our uncertainty contour. Admittedly, we do not get a very tight constraint: the contour that we draw is $\sim 20\degree$ long in its elongated direction. However, only a few works \citep[e.g.,][]{erkal2019} have reported the uncertainties on this quantity. This does not come as a surprise since the simulated stream is formed and evolved inside the full cosmological volume, for which we have assumed a quite simplistic potential model. The stream stars have also been stripped at several different pericentric passages and may or may not have close encounters with subhalos. Moreover, the assumptions and approximations in our numerical methods can further introduce another layer of uncertainties (\S\ref{sec:degeneracies} and \S\ref{sec:m12i_uncertainties}). Given these assumptions, it is surprising that this method still constrains the DM halo minor axis to the extent that we have shown.

To summarize, we test the method described in \S\ref{sec:algo} of constraining the DM halo symmetry axis on a simulated tidal stream which is considered to be a Sgr stream analog. We sample trial rotations in Galactic latitude and longitude throughout the entire northern hemisphere. Assuming an axisymmetric multipole potential, we compute the slope difference $\Delta$ at each trial rotation and find a region that yields minimal $\Delta$ (Figure \ref{fig:sim_slope_diff}). We then set up a zoom-in grid of trial rotations around this region and repeat our procedures. Finally, we draw our uncertainty contour containing the predicted location of the DM halo minor axis, which is in agreement with the true DM halo minor axis location (\ref{fig:m12i_uncertainty}).

\section{the Sagittarius stream}\label{sec:sag_data}
In this section, we apply the method to the Sgr stream to constrain the orientation of the MW's DM halo.

\subsection{data}

\begin{figure*}
\plotone{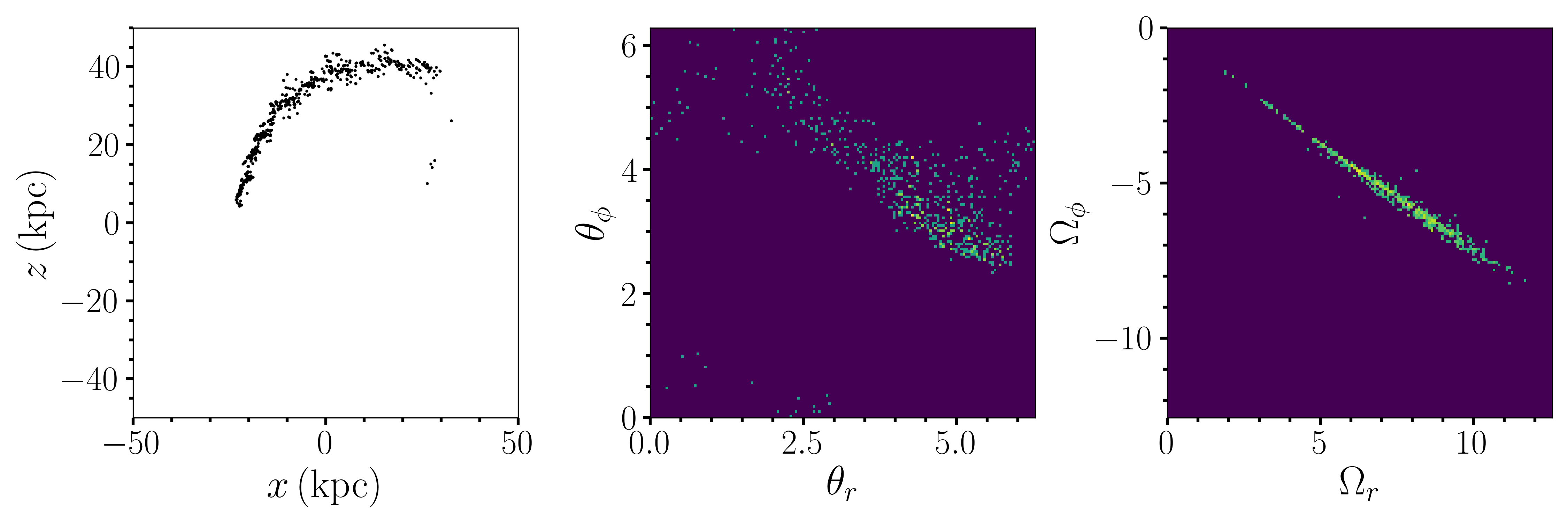}
\caption{Different 2D projections of the leading arm of the Sgr stream that we use to constrain the potential of the MW. The projections are real space $x-z$ projection (left), angle space $\theta_r-\theta_\phi$ projection (middle) and frequency space $\Omega_r-\Omega_\phi$ projection (right). Everything is computed and plotted in the default host principal axis frame at the present day, where the $z$ direction points perpendicular to the disk. The assumed potential is described in \S\ref{sec:pot2}, and we set the halo flattening parameter $q=0.8$. \label{fig:sgr_stream}}
\end{figure*}

\cite{vasiliev2021} identified 4465 Sgr stream candidate members with 6D phase-space information using data from surveys and catalogues such as \emph{Gaia} DR2 \citep{gaia-dr2}, 2MASS \citep{2mass}, APOGEE \citep{apogee}, LAMOST\citep{lamost}, \emph{Gaia} RVS \citep{gaia-rvs}, Simbad and \cite{2011ApJ...727L...2P}. These stars are divided into the leading arm in the northern Galactic hemisphere and the trailing arm in the southern hemisphere. The area in between the two arms is the Galactic plane which obscures stream candidate member identification as well as the surviving bound part of the Sgr stream. Most of these candidates are in the Red Giant Branch (RGB) since these are the brightest, yielding the lowest proper motion uncertainties. \cite{vasiliev2021} first performed a cross-match between Gaia DR2 and 2MASS and selected stars that are within their desired ranges of Galactic latitude, G magnitude, BP-RP, and parallax. The likely stream candidates were then selected from this sample from the cuts in proper motion, Gaia color-magnitude, and 2MASS color. The heliocentric distance is assigned to the RGB stream candidates as the median of the 5 nearest RR Lyrae stars on the sky. See their Section 2.3 for more detail on how these stream candidates were identified. We have observational uncertainties for 4 of the quantities, while the declination and right ascension have negligible uncertainties.

\subsection{potential model}\label{sec:pot2}

We use s simple 3-component potential similar to \texttt{MWPotential2014} presented in \texttt{galpy} \citep{bovy2015}. The bulge, disk, and halo are represented by a spheroid profile, Miyamoto-Nagai potential \citep{1975PASJ...27..533M}, and flattened spheroid profile, respectively. Since we are not focusing on the overall shape of the potential, we keep most the parameters fixed, except the flattening of the halo component in the $z$ direction ($q<1$ implies oblate). The spheroid profile is given by
\begin{equation}\label{eq:spheroid}
    \rho = \rho_0 \left(\frac{\tilde{r}}{r_s}\right)^{-a}\left(1 + \frac{\tilde{r}}{r_s}\right)^{-b} \times \,\mathrm{exp}\left[-\left(\frac{\tilde{r}}{r_{cut}}\right)^2\right],
\end{equation}
where $\tilde{r} = \sqrt{x^2 + (y/p)^2 + (z/q)^2}$ is the ellipsoidal radius. The other parameters are the density norm $\rho_0$, scale radius $r_s$, cutoff radius $r_{cut}$, exponents ($a,b$) and flattening parameters ($p,q$). The Miyamoto-Nagai potential is
\begin{equation}\label{eq:miyamoto}
    \Phi = \frac{M}{\sqrt{R^2 + \left(r_s+\sqrt{z^2 + r_h^2}\right)^2}},
\end{equation}
where $R = \sqrt{x^2 + y^2}$ is the cylindrical radius. The other parameters are mass $M$, scale radius $r_s$ and scale height $r_h$. All of the parameters were picked such that the circular velocity is $\sim 220$ km/s at the solar position of 8 kpc. The summary of all the parameters is shown in Table \ref{table:potentials}.

\begin{deluxetable*}{lccccccccc}
\tablehead{
\colhead{component} & \colhead{$\rho_0$} & \colhead{$M [\Msun]$} & \colhead{$r_s$[kpc]} & \colhead{$r_h$[kpc]} & \colhead{$r_{cut}$[kpc]} & \colhead{a} & \colhead{b} & \colhead{p} & \colhead{q}}
\startdata
bulge & $2.23\times10^{8}$ & - & 1 & - & 1.9 & 1.8 & 0 & 1 & 1\\
disk & - & $6.82\times10^{10}$ & 3 & 0.28 & - & - & - & - & -\\
halo & $8.49\times10^{6}$ & - & 16 & - & 0 & 1 & 2 & 1 & varied\\
\enddata
\caption{The MW potential consists of the bulge, disk, and halo. The bulge and halo follow the spheroid profile. The disk follows Miyamoto-Nagai potential. The analytical forms for the profile and potential, which explain the meaning of each parameter in this table, are given in Equation \ref{eq:spheroid} and \ref{eq:miyamoto}. \label{table:potentials}}
\end{deluxetable*}

Assuming this axisymmetric potential, we can then estimate the action, angle, and frequency variables of the Sgr stream members that have 6-D phase-space information using \code{AGAMA}. The frequency variables still look coherent and spread into a line structure as we expected. However, the structures in the angle space are more diffuse, compared to the simulated stream. The trailing arm is much more heavily impacted as we can hardly notice any line-like structure. This could be due to the location of the trailing arm which lies in the southern hemisphere closer to the LMC. Stream stars in other streams are shown to exhibit the misalignment between their velocities and stream track where they pass through the southern hemisphere  \citep{shipp2019,li2021}, pointing towards the past locations of the LMC. We consider the decoherence of the angle-frequency space of the trailing arm of the Sgr stream to be additional evidence of this type of interaction since the stream track-velocity misalignment seen in these other streams is simply the real-space reflection of a severe angle-frequency misalignment. For our purpose, we discard the trailing arm and only use the leading arm (613 stars) in this work. Figure \ref{fig:sgr_stream} shows the leading arm of the Sgr stream in real space $x-z$ projection (left), angle space $\theta_r-\theta_\phi$ projection (middle) and frequency space $\Omega_r-\Omega_\phi$ projection (right). These are shown in the default Galactocentric frame, where the $z$ direction points perpendicular to the disk. We note the similarities between Figure \ref{fig:sim_stream} and \ref{fig:sgr_stream}. The two main differences are the length of the Sgr stream leading arm in which we have complete 6D phase-space information and the rotational sense of the orbit in this particular rotation frame, as shown by the negative $\Omega_\phi$ value in the right panel. Because of its shorter length, the Sgr stream does not suffer from the degeneracies described in \S\ref{sec:degeneracies} for the simulated stream.

\subsection{the halo flattening parameter q}\label{sec:q}

\begin{figure}
\plotone{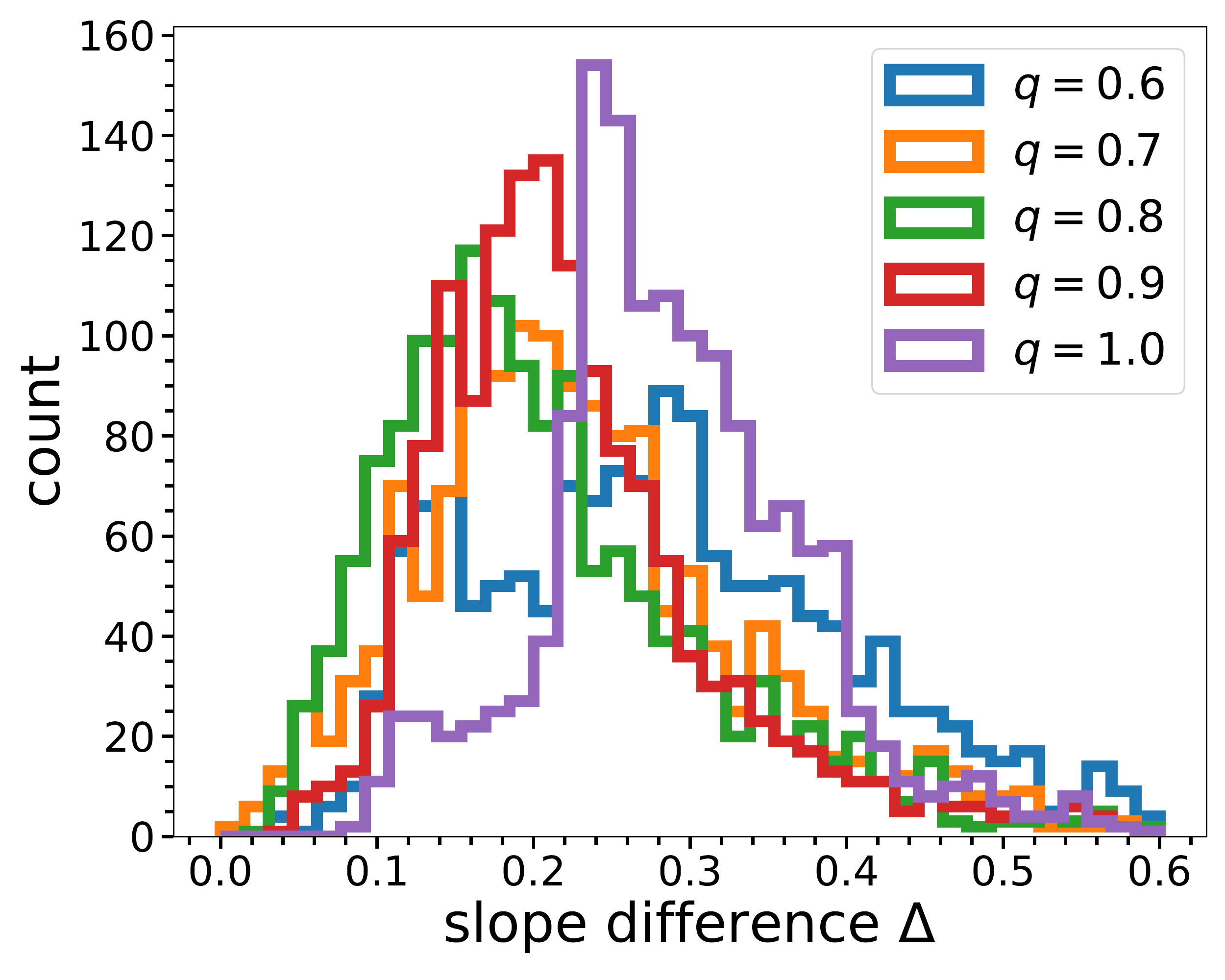}
\caption{Distribution of slope difference $\Delta$ obtained with the leading arm of the Sgr stream for different values of halo flattening parameter $q\in \{0.6, 0.7, 0.8, 0.9, 1.0\}$, where $q=1$ is spherical and $q<1$ is oblate. For each $q$, we estimate $\Delta$ at a set of trial rotations, spreading across the northern hemisphere in Galactic coordinate. Each color represents a fixed $q$. \label{fig:constrain_q}}
\end{figure}

To quantify the effect of the flattening parameter on our results, we test our method on the Sgr stream leading arm using the potential described in \S\ref{sec:pot2} for $q \in \{0.6, 0.7, 0.8, 0.9, 1.0\}$, ranging from more oblate ($q<1$) to perfectly spherical ($q=1$). We do not try $q > 1$ since the Stack\"el fudge approximation \citep{2012MNRAS.426.1324B} only works for disk-like (oblate) potentials; however, the range of $q$ for oblate potentials already brackets the smallest minimum $\Delta$, so we do not expect that prolate potentials would improve matters. In this subsection, we ignore measurement uncertainties. We attribute this effect to the observational uncertainties and the LMC, which will be discussed in detail in \S\ref{sec:discussion}. For each $q$, we use the same set of trial rotations as in the simulated stream case described in \S\ref{sec:results_sim} and estimate the slope difference $\Delta$ at each trial rotation. Figure \ref{fig:constrain_q} shows the distributions of the estimated $\Delta$ at different values of $q$, where each $q$ is shown with a different color. Overall, the estimated $\Delta$ is roughly an order of magnitude larger than what we see in the simulated stream case. This seems sensible since the line in the angle space (in Figure \ref{fig:sgr_stream}) is much more diffused (compared to Figure \ref{fig:sim_stream}). For intermediate values $q \in \{0.7, 0.8, 0.9\}$, the distributions are somewhat similar as they peak around $\Delta\sim 0.2$ and are able to achieve low $\Delta$, compared to when we use a more extreme $q$ of either 0.5 or 1.0. The regions that minimize $q$ are not identical for every value but are within the same octant on the sphere. $q=0.7$ (orange line) is able to achieve the lowest $\Delta$. However, it's difficult to argue that $q=0.7$ is a better fit than $q=0.8-0.9$ since the number of trial rotations with these lowest $\Delta$ are small. $q=0.8$ (green line), on the other hand, has the lowest peak value of $\Delta$ and seems to produce the largest number of trial rotations with low $\Delta$. On the extreme ends, the more oblate and the more spherical halos yield objectively worse fits as their distributions peak around $\Delta\sim 0.3$. There are significant overabundances of $q=0.6$ and $q=1.0$ at $\Delta > 0.25$. Since our goal is not to study the halo shape, \textbf{we put a weak constraint on the halo flattening parameter at $q=0.7-0.9$} and proceed with modeling the halo tilt using the value $q=0.72$, which is consistent with the above findings. It is also consistent with the halo shape reported in \cite{law2010}, which uses Sgr to constrain the MW halo potential with a different technique. Despite their model being triaxial, it is very close to oblate, reporting $b/a=0.99$ and $c/a=0.72$.


\subsection{Results}

\begin{figure}
\plotone{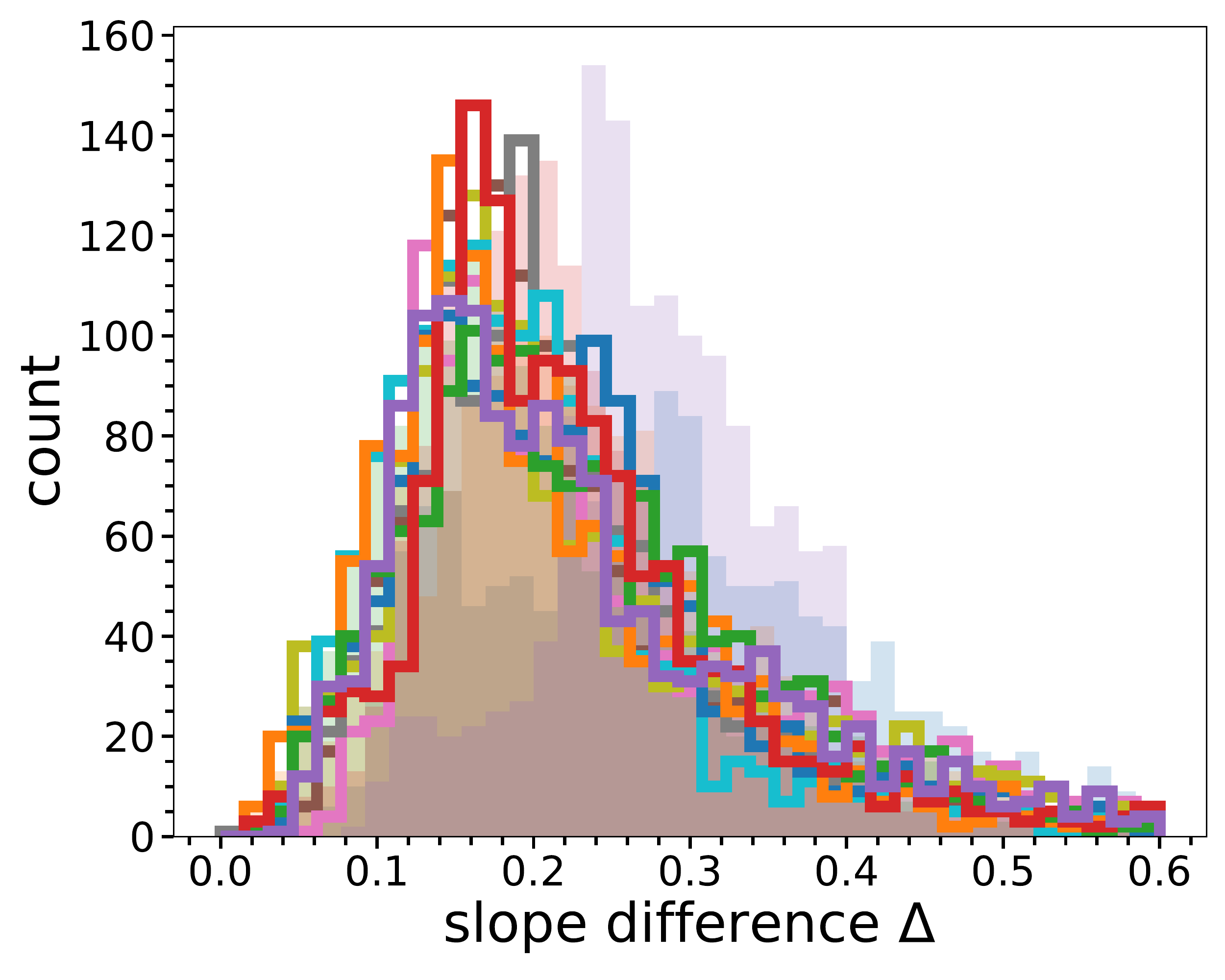}
\caption{Similar to Figure \ref{fig:constrain_q}, we compare the distributions of $\Delta$ with $q=0.72$ for all 10 independent realizations of the Sgr stream's leading arm data with observational uncertainties. Each realization is shown by a unique color. The shaded regions in the background are the histograms from Figure \ref{fig:constrain_q}. We show that $\Delta$ is more sensitive to the flattening parameter $q$ than the observational uncertainties. \label{fig:compare_realization}}
\end{figure}

\begin{figure*}[t]
\plotone{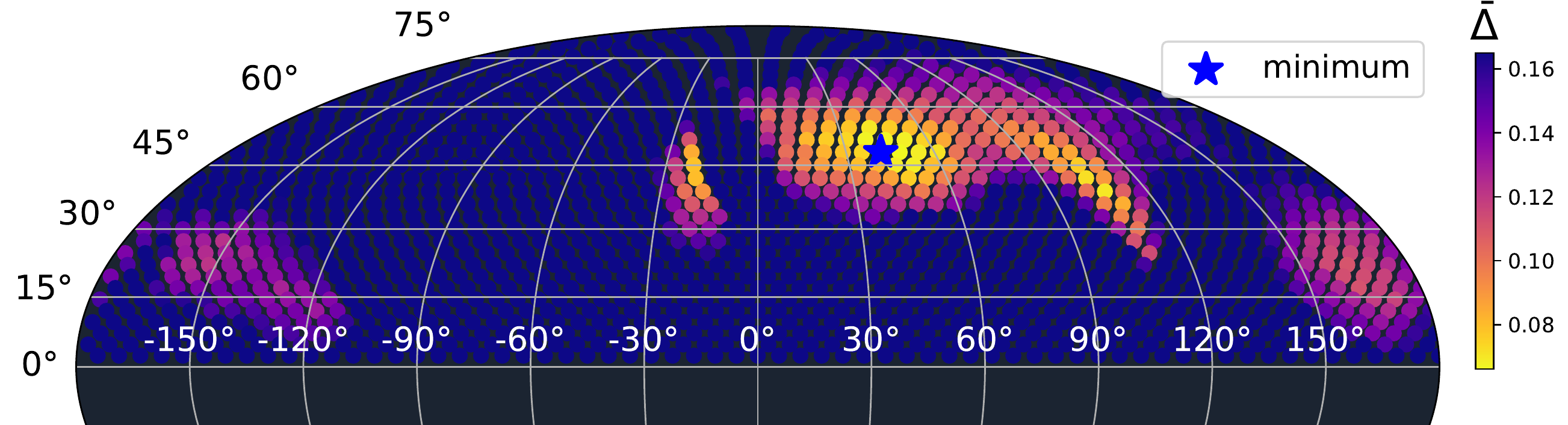}
\caption{Results for the MW halo's tilt from fits to the leading arm of Sgr stream. To incorporate the observational uncertainties, we compute the average slope difference $\bar{\Delta}$ across 10 independent realizations of the leading arm for each trial Galactic latitude and longitude location. The angle and frequency variables are computed by assuming the potential described in \S\ref{sec:pot2} with $q=0.72$ and various trial values of $(\ell, b)$. All the trial rotations are shown with filled circular markers, with the color representing estimated $\bar{\Delta}$. The blue star marks the global minimum. \label{fig:constrain_axis}}
\end{figure*}

\begin{figure*}[t]
\plotone{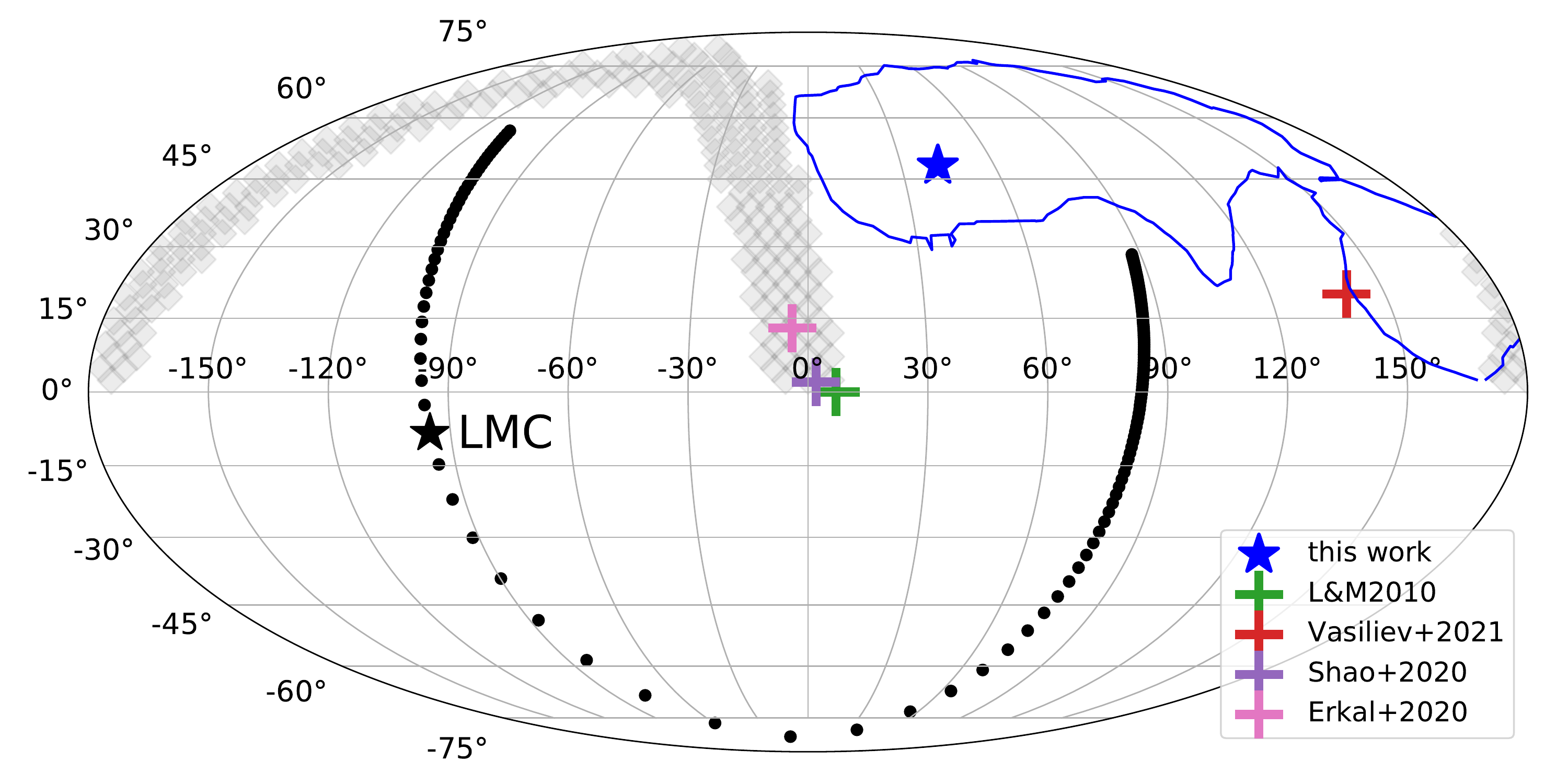}
\caption{Results for the leading arm of Sgr stream. We find the best-fit location for the MW's DM halo minor axis at $(\ell,b)=(42\degree,48\degree)$ (shown by the blue star). Its associated uncertainty is shown by the contour of the same color. For comparison, we show the estimates from other works in crosses with different colors. We identify the locations associated with the failure mode in gray. The estimated LMC orbital trajectory and its present-day location from \cite{garavito2019}'s simulation are shown by the black markers and the black star, respectively. \label{fig:sgr_uncertainty}}
\end{figure*}

Unlike the simulated stream, the Sgr stream data contain observational uncertainties. To account for the observational uncertainties for each stream star, we draw its proper motions, distance, and radial velocity from normal distributions with the means centered at the observed values and the standard deviations being their associated uncertainties. We treat uncertainties in declination and right ascension as negligible and use the same value across all realizations. Next, we construct 10 independent realizations of the observed Sgr streams (by drawing the values for each star 10 times). For each realization, we repeat the procedure in \S\ref{sec:q}, but with $q$ fixed at $q = 0.72$, and estimate the slope difference $\Delta$ and its uncertainty $\sigma_\Delta$ at each trial rotation. Looking at each realization independently, we see that the minimum regions are generally similar, but the overall $\Delta_{min}$ vary a little. In Figure \ref{fig:compare_realization}, we compare the distributions of $\Delta$ with $q=0.72$ for all 10 independent realizations of the Sgr stream's leading arm data with incorporated observational uncertainties. Each realization is shown by a unique color. We also show the shaded regions in the background, which are the histograms from Figure \ref{fig:constrain_q}. At fixed $q$, the distribution for each realization displays a similar mode and spread compared to the shaded histograms in the background. We conclude that $\Delta$ is more sensitive to the flattening parameter $q$ than the observational uncertainties.  For the $i^{th}$ realization, we denote its slope difference at location $(l, b)$ as $\Delta_i(l, b)$. At each trial rotation, we compute the average slope difference,
\begin{equation}
    \bar{\Delta}(\ell,b) = \frac{1}{10}\sum_{i=1}^{10} \Delta_i(l, b),
\end{equation}
and the average uncertainty,
\begin{equation}
    \sigma_{\bar{\Delta}}(\ell,b) = \frac{1}{10}\sqrt{\sum_{i=1}^{10}  \sigma^2_{\Delta_i}(\ell,b)}.
\end{equation}
The results are shown in Figure \ref{fig:constrain_axis}. Similar to Figure \ref{fig:sim_slope_diff}, the filled circular markers show all the trial rotations. Their colors show the estimated average slope difference $\bar{\Delta}$ computed across 10 independent realizations of the leading arm. The blue star marks the location of the global minimum at $(\ell,b)=(42\degree,48\degree)$. The shape of the region surrounding the minimum is peculiar as there is a band of poorly fitted regions going across it to the left (around $(\ell,b)=(-10\degree,45\degree)$). This band corresponds to a failure mode of this method, which is discussed in more detail in \S\ref{sec:discussion}. With the average uncertainties $\sigma_{\bar{\Delta}}$, we inspect the uncertainty estimates around the minimum point and confirm that the values are smoothly transitioning. Following Equation \ref{eq:uncertainty_countour}, we choose to draw the density contour around the region that satisfies $\bar{\Delta} < \bar{\Delta}_{min} +\sigma_{\bar{\Delta}_{min}}$. We find $\bar{\Delta}_{min} = 0.066$ and $\sigma_{\bar{\Delta}_{min}} = 0.101$, hence our uncertainty contour is described by  $\bar{\Delta} < 0.167$. Figure \ref{fig:sgr_uncertainty} shows this uncertainty contour in blue. Since  $\sigma_{\bar{\Delta}_{min}}$ is quite large, the contour that we draw, and hence the constraint on the MW halo symmetry axis, is weak. We also plot the estimated minor axis location from other works in the literature. These include the estimations from \cite{law2010} (green; hereafter L\&M2010), \cite{vasiliev2021} (red; hereafter Vasiliev+2021), \cite{shao2020} (purple; hereafter Shao+2020) and \cite{erkal2019} (pink; hereafter Erkal+2020). L\&M2010 perform orbit fitting of Sgr stream using a triaxial potential, finding nearly oblate DM halo shape with the minor axis at $(\ell,b) = (7\degree,0\degree)$. Vasiliev+2021 perform a more robust fitting of the Sgr stream considering the presence of the LMC and constrains the minor axis of the outer halo at $(\ell,b) = (140\degree, 20\degree)$. Shao+2020 infers the minor axis from the co-planar orbits of the MW's satellites, reporting the halo minor axis at $(\ell,b) = (2\degree, 2\degree)$. Lastly, Erkal+2019 reports the minor axis at $(\ell,b) = (-4\degree, 13\degree)$ by modeling the velocity misalignment along the Orphan stream in the presence of the LMC with an oblate halo and a reflexive model where the MW can move in response to the LMC. \textbf{Despite the large uncertainty contour, our results are inconsistent with the estimates from these reported values}. These inconsistencies may be partially explained by the locations of the failure mode and the effects of the LMC, as discussed in \S\ref{sec:discussion}. We show the failure mode locations in gray, which incldue the MW's DM halo minor axis estimates from L\&M2010, Shao+2020, and Erkal+2020. We also show the estimated locations of the past and future LMC orbit from \cite{garavito2019}'s simulation in black. The black star marks the present-day location of the LMC in the simulation. The black scatter points are plotted every 20 Myr, starting from $\sim2$ Gyr ago before the present day. In this projection, the LMC is travelling from right ($l > 0\degree$) to left ($l < 0\degree$).

\section{Discussion}\label{sec:discussion}
Using the leading arm of the Sgr stream, we cannot put as tight of a constraint on the location of the DM halo minor axis, compared to the simulated stream test case. We report that MW's DM halo is oblate with $q\sim 0.7-0.9$. However, Figure \ref{fig:sgr_uncertainty} shows the tension between our prediction and other estimates of the MW's DM halo minor axis location. In fact, the results are not surprising since the observed data are not error-free and the MW is undergoing a merger with the LMC, its most massive satellite. However, it is ambiguous how both of these affect our angle and frequency computation, our method, and our predictions since the Sgr data is limited to the present day. To quantify these effects, a comprehensive study involving the LMC-analog and tidal streams of different progenitor masses, orbits, and disruption times will need to be done. This is beyond the scope of this manuscript since our goal is to test and identify limitations of the method first proposed in \cite{sanders2013a}, which is slightly modified and described in \S\ref{sec:method}. We believe that systematic uncertainties from the Stack\"el fudge approximation are not enough to explain the discrepancies since they should have higher influences on the stream at the tail ends \cite{sanders2013a} on the computed angles and frequencies. However, we only incorporate the middle sections of the stream in both the simulated stream (where we only use the most prominent line detected) and the Sgr stream (where we only use the leading arm). For the rest of this section, we explore other possible explanations for these inconsistencies.

\subsection{radial dependence of the halo shape and orientation}
The inconsistencies alone do not invalidate our results. Cosmological simulations of galaxy formation have shown that it is common for the DM halo to exhibit a radius- and time-dependent large-scale tilt \citep[e.g.][Baptista et al. in prep.]{vasiliev2021,emami2021}. Because of the radial dependence, it is not sufficient to describe the disk-halo misalignment with a single parameter. Using 25 MW-like galaxies from the \code{Illustris TNG50} simulation, \cite{emami2021} uses the enclosed volume iterative method to measure the DM halo shape at different radii. They show that the majority of the DM halos experience either gradual or abrupt rotations throughout their radial profiles. Baptista et al. in prep. also find similar results in simulated MW-mass hosts from a suite of FIRE simulations \citep{Wetzel2016} which appear to be linked to their formation histories. \cite{vasiliev2021} model the full interaction between the MW, LMC, and Sgr. They report their models generally produce an oblate halo that aligns with the disk in the inner part while becoming triaxial and misaligned with the disk in the outer part beyond $\sim50$ kpc. Hence, the predictions of the tilt of the MW's DM halo and the minor axis location reported by previous studies might probe different radial ranges and times based on the methods employed. L\&M2010 use a static external MW potential with a fixed orientation with a triaxial shape. Erkal+2019 use the Orphan stream which probes the inner region of the halo at $r_{gal} \gtrsim 10$ kpc. Shao+2020 only reports a fixed MW halo minor axis using the argument on the plane of satellites. Finally, the prediction from Vasiliev+2021 is probing the \emph{outer} halo. There are other estimates of the MW halo minor axis location that we do not plot in Figure \ref{fig:sgr_uncertainty}, mostly due to them not reporting explicit values. For example, \cite{han2022} argues that the MW's DM halo is oblate and is tilted $\sim30\degrees$ above the Galactic disk towards the Virgo Overdensity in the North and the Hercules-Aquila Cloud in the South. Using a simulated stream, we have shown in \S\ref{sec:sim_data} that our method predicts the minor axis location that is consistent with the minor axis of the DM halo at $\sim50$ kpc, similar to the galactocentric distances of the stream stars. Since the simulated stream stars are orbiting in not very eccentric orbits, it is possible that radial streams will probe different radial ranges that differ from their present-day positions. Our Sgr stream's leading arm data shows the average distance of $\sim35$ kpc, which should be similar to the radial scales of the results that we report in \S\ref{sec:sag_data}. Moreover, \cite{buist2017} demonstrates that the behavior of a stream in the angle and frequency spaces is sensitive to the time-dependent aspect of the potential, suggesting that our method might be more sensitive to the \emph{past} orientation of the DM halo (e.g. pre-LMC infall), rather than its present-day tilt.

\subsection{observational uncertainties as the main source of the failure mode}

\begin{figure}[t]
\plotone{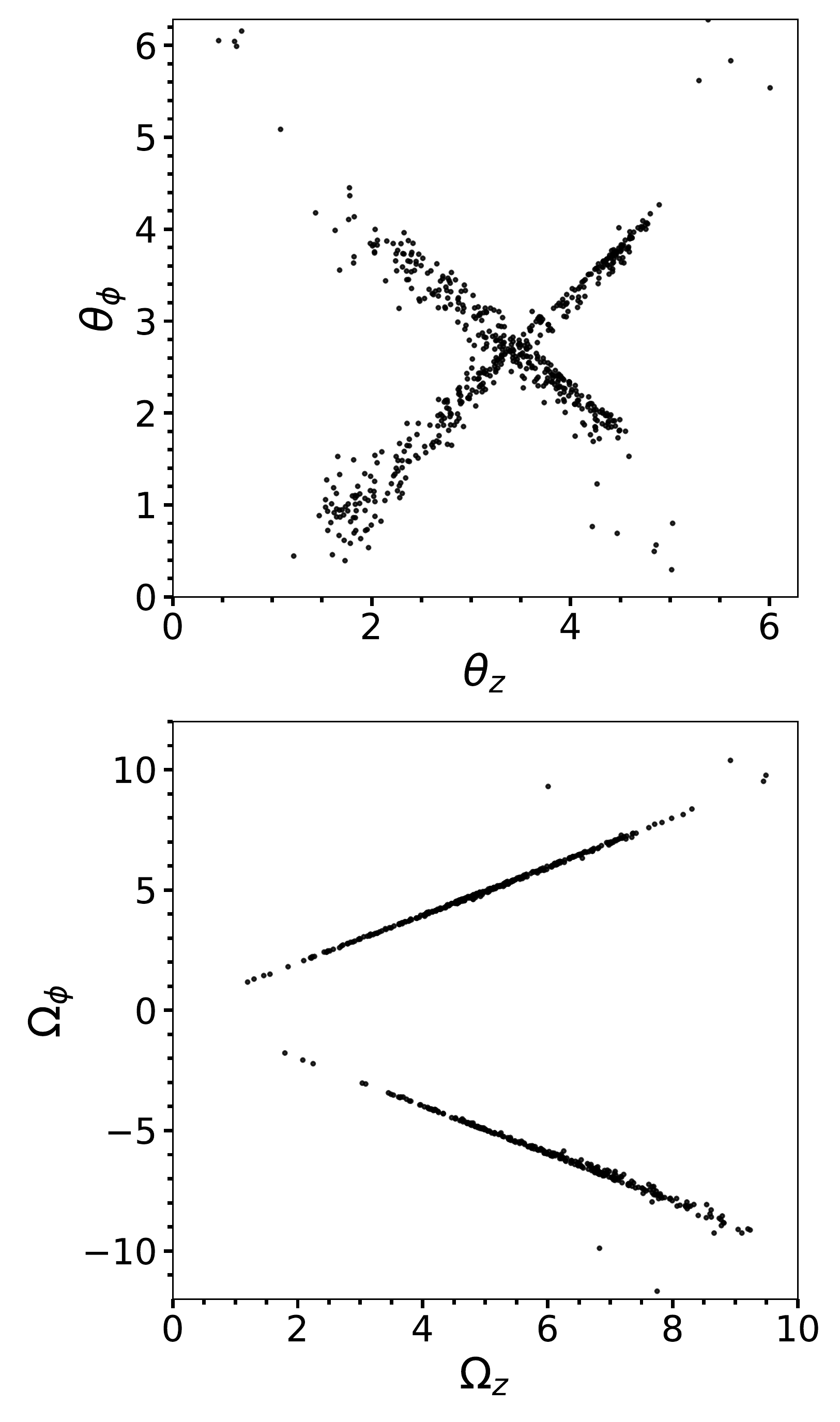}
\caption{An example of the failure mode in which stream stars are splitting into both prograde ($\Omega_\phi > 0$) and retrograde orbits ($\Omega_\phi<0$) at certain trial rotations shown by the gray band in Figure \ref{fig:sgr_uncertainty}. Sgr stream stars in the leading arm (without uncertainties) are shown in 2-D angle projection (top) and 2-D frequency projection (bottom) corresponding to $(\ell_{trial}, b_{trial})=(-10\degree,45\degree)$. \label{fig:failure_mode}}
\end{figure}

In Figure \ref{fig:sgr_uncertainty} we show a band of poorly-fitted regions identified by our method in gray color. We refer to these regions as the failure mode and argue that one of the main sources of the failure mode is observational uncertainties. The failure mode arises due to the fact that a stream does not delineate a single orbit \citep{sanders2013b}. The orbits of the stream stars do not share the exact same plane as their orbital planes generally spread around the common average plane. Therefore, if this average orbital plane lies close to the z-axis in an arbitrary coordinate system, some stars will have prograde orbits, while others have retrograde orbits. This results in stream splitting in angle and frequency space where the prograde stars have $\Omega_\phi > 0$, while the retrograde stars have $\Omega_\phi < 0$. The gray band in Figure \ref{fig:sgr_uncertainty} shows the trial rotations $(\ell_{trial}, b_{trial})$ that exhibit the failure mode. Specifically, we refer to a failure mode when at least 20 percent of the stream stars have opposite $\Omega_\phi$. Figure \ref{fig:failure_mode} shows an example of the failure mode corresponding to $(\ell_{trial}, b_{trial}) = (-10\degree,45\degree)$. The stars belonging to the Sgr stream's leading arm are shown in the $\theta_z-\theta_\phi$ (top) and $\Omega_z-\Omega_\phi$ (bottom). The failure mode can directly affect our estimation of the slope difference $\Delta$. If the splitting is almost equal, the implemented line detection algorithm might incorrectly identify the appropriate lines, yielding one line with a positive slope and the other with a negative slope. This results in the outliers in $n_{r\phi}/m_{r\phi}$ and $n_{z\phi}/m_{z\phi}$, hence, the overall poor-fitting for $\Delta$ following Equation \ref{eq:slope_diff}. Moreover, the crossing of the lines in angle space can bias the linear-regression fit even if the algorithm detects appropriate lines.

One of the main sources of the failure mode is the observational uncertainties in the Sgr stream data. The orbit of any given star is dictated by its 3-D position and its 3-D velocity, which can be biased by observational uncertainties. The Sgr stream leading arm, despite being much shorter than the simulated stream considered in \S\ref{sec:sim_data}, displays a much greater spread in the orbital planes. The simulated stream data which is error-free also exhibit a failure mode, yet its appearance is negligible, impacting only a few trial rotations. Since the failure mode contains the MW's DM halo minor axis predictions from L\&M2010, Shao+2020, and Erkal+2020, we cannot accurately estimate $\Delta$ at those locations. \cite{sanders2013a} suggests binning as a way to reduce observational uncertainties. However, since we only have 613 stars in the Sgr stream's leading arm, we find this approach impractical since we need $\sim 500$ stars to apply our method.

\subsection{the LMC}
The LMC is the most massive satellite of the MW. It is currently just past its first pericentric passage in a most likely very eccentric orbit \citep{besla2007, kallivayalil2013}. It has been studied in tailored N-body simulations that such an interaction with a massive satellite can induce both \emph{transient} and \emph{collective} responses within the host, affecting the densities and kinematics of the host's DM halo \citep{garavito2019}. The transient response or ``wake'' is an overdensity tracing the past orbit of the LMC and is predicted to persist in the radial range $r_{gal} = 45-200$ kpc. In contrast, the collective response or ``reflex motion'' is generated by resonant displacement of the orbital barycenter. This creates an overdensity region in the north extending to all distances.  These two responses can also be observed in the host's stellar halo \citep{cunningham2020, conroy2021, petersen2021}.

\begin{figure}[t]
\plotone{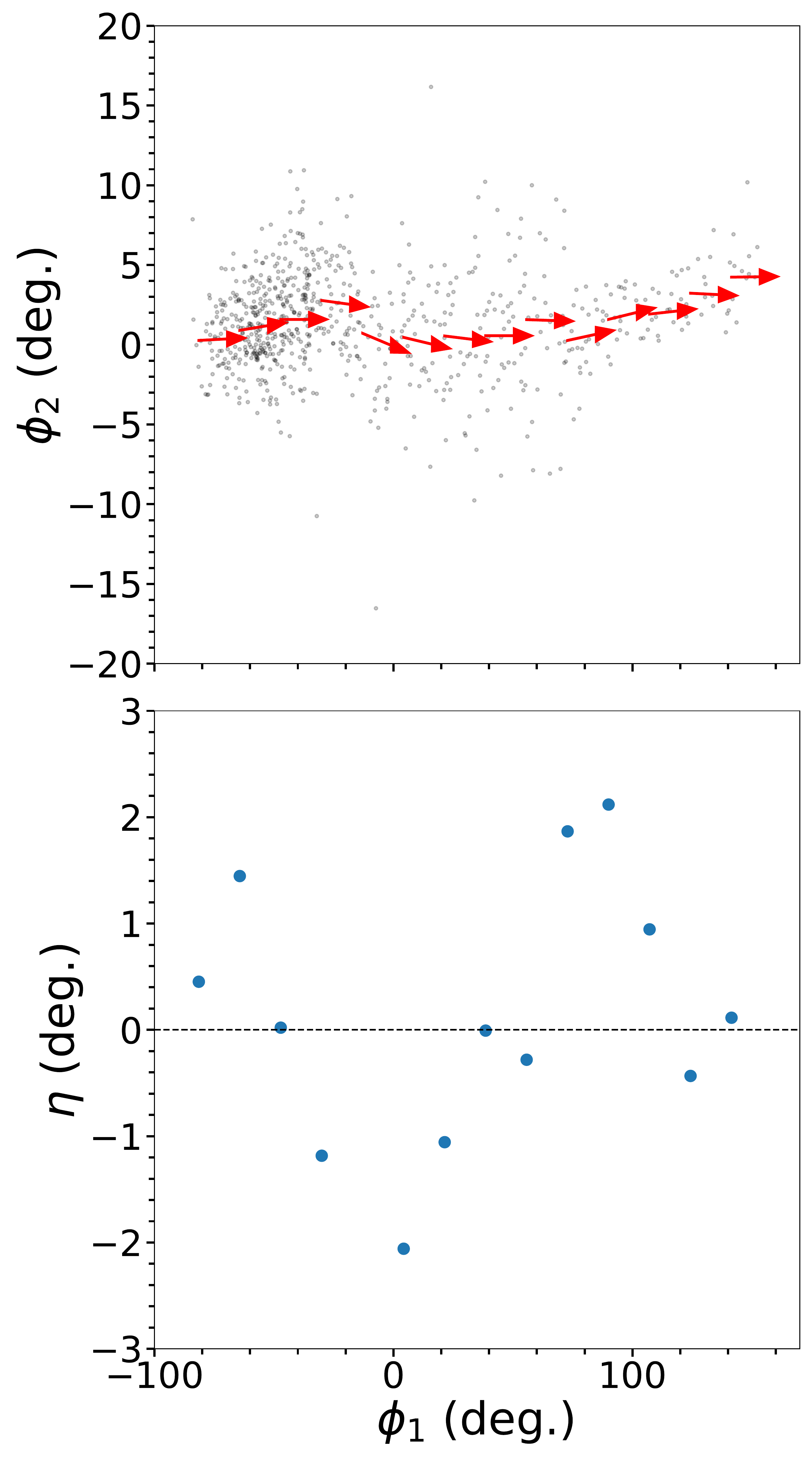}
\caption{Top: the middle section of the simulated stream from \S\ref{sec:sim_data} shown in the stream coordinate ($\phi_1,\phi_2$). The stream stars are binned along $\phi_1$. The red arrows show the normalized direction of the mean velocity of the stream stars in each bin (reflex-corrected). Bottom: the offset angle $\eta$ between the mean velocity vector and the $\phi_1$ axis defined as $\phi_2 = 0\degree$. We show that the simulated stream does not exhibit stream velocity misalignment. \label{fig:m12i_vel_alignment}}
\end{figure}

The extent to which the LMC affects the Sgr stream's leading arm (specifically in the angle and frequency space) is not well-understood. The stars in the leading arm are at distances $r_{gal} \sim 35\pm10$ kpc, right in the transitioning distance scales where we expect stronger perturbations from the LMC. On one hand, we expect stars with small orbital timescales (smaller than the timescale in which the MW moves by a distance equal to the barycenter displacement) to respond adiabatically to the LMC's passage. These stars will move along with the reflex velocity of the MW. \cite{erkal2019} estimates this corresponds to the region within $r_{gal}<30$ kpc, hence the stars at these small distance scales should conserve their actions as well as the features in angle and frequency space, suggesting that our method might be more sensitive to the past orientation of the MW's DM halo (i.e. before LMC-infall). On the other hand, \cite{garavito2019} reports perturbations in the density field of the MW's DM halo are largest at $r_{gal} \gtrsim 45$ kpc, reflecting the time the LMC has spent in the outer regions. The wake is weak at $r_{gal} < 45$ kpc, yet the LMC can still affect the outer disk \citep{laporte2018}. We plot the estimated LMC orbital trajectory and its present-day location from \cite{garavito2019}'s simulation in Figure \ref{fig:sgr_uncertainty} using black markers. In this projection, the LMC enters the MW's virial radius $\sim 2$ Gyr ago from the right $l\sim90\degree$. Interestingly, in Figure \ref{fig:constrain_axis} we show a strange well-fitted region extending vertically between $90\degree < l < 120\degree$. This vertical streak coincides with the LMC's passage when it first enters the halo. It is suggestive that the LMC has an impact on the method and can possibly affect our results. 

The LMC can also induce misalignment between the velocity vector and stream track \citep{shipp2019, shipp2021, li2021}. These velocity misalignments cannot be induced from a static external MW potential and have been observed in several streams \citep{erkal2019, vasiliev2021}. Despite being in the northern hemisphere, the leading arm of the Sgr stream still exhibits the velocity misalignment \citep[see Figure 2 in][]{vasiliev2021}. In contrast, the host in the simulation $\texttt{m12i}$ is much more dormant and has not undergone any major majors within the past $>5$ Gyr. To quantify the stream track--velocity misalignment in the simulated stream, we first identify the middle section of the stream corresponding to the most prominent line in the angle space in the default galactocentric coordinate frame. We then transform the stream into the heliocentric coordinate system using \code{Astropy} \citep{astropy:2013,astropy:2018,astropy:2022}. We fit a great circle through the stream and transform the data into the stream coordinate $(\phi_1,\phi_2)$, where $\phi_1$ is the direction along the stream and the spread in $\phi_2$ measures its thickness. The proper motions are solar reflex-corrected. Figure \ref{fig:m12i_vel_alignment} shows the simulated stream in the stream coordinate. In the top panel, the stream stars are binned along $\phi_1$. The direction of the mean velocity in each bin is shown by the red arrow (normalized). In the bottom panel, we compute the offset angle $\eta$ between all the red arrows and the $\phi_1$ axis (defined by $\phi_2=0\degree$) and we plot them against the bin centers along $\phi_1$. Overall, we find that the simulated stream does not suffer from stream velocity misalignment, with $\eta<3\degree$ in all bins. This is suggestive that the degree of stream track-velocity misalignment can be used to predict how well our method will perform, with the best results coming from streams that exhibit minimal misalignment.

\section{Summary}\label{sec:summary}
In this manuscript, we revisit the method of constraining the Galactic potential using tidal streams in angle and frequency space first introduced in \cite{sanders2013a}. Under the correct potential, we expect the slopes of the lines in angle and frequency space to best match up as the stream should grow in the direction parallel to $\hat{e}_1$, a principal eigenvector of the Hessian with the largest eigenvalue, in both angle and frequency. \cite{sanders2013a} tests this method on a simulation of a cold, thin stream evolving under an external static MW potential of logarithmic form with 2 free parameters. They show that they can correctly recover the true parameters with this method. However, the real MW potential is a complex, clumpy, and time-evolving potential. The real streams evolving inside the MW are susceptible to satellite and dark subhalo interactions, non-adiabatic evolution, and resonant trapping, producing observational features like spurs, gaps, and bifurcations \citep[e.g.][]{carlberg2013,pearon2017,bonanca2019,tomer}. These can cause the behavior of the real streams to deviate from the theoretical expectations, especially in the action-angle framework \citep{10.1046/j.1365-8711.1999.02616.x, 10.1046/j.1365-8711.1999.02690.x}. We thought this method should be tested under a more realistic setup. 

We extend this method and test it on a simulated (from the full cosmological simulation of galaxy formation) and real dwarf galaxy stream. Because of the difficulty of expressing the realistic MW potential in the analytical form, our goal is to constrain the tilt of the DM halo (specifically by constraining the location of the minor axis) by assuming oblate axisymmetric potentials. We define the goodness-of-fit parameter to be the slope difference $\Delta$ described by Equation \ref{eq:slope_diff}. The followings are the key results from this manuscript.

\begin{enumerate}
    \item Assuming a low-order axisymmetric multipole potential, we show that a hot and diffuse dwarf galaxy stream simulated under a realistic environment of the full cosmological simulation still behaves according to the theoretical expectations in the action-angle framework. We successfully \emph{unwrap} the long, convoluted stream in Figure \ref{fig:unwrap}. Moreover, a stream in angle and frequency space is less sensitive to the overall shape of the potential but more sensitive to the symmetry axis (from the results shown in \S\ref{sec:sim_data}).
    \item Using the simulated stream with error-free measurements and a quiet host galaxy, we can successfully recover the minor axis location of the DM halo at the distance scales similar to the distances of the stream stars.
    \item Using the Sgr stream's leading arm, we put weak constraints on the halo flattening parameter $q\sim0.7-0.9$ and its minor axis location $(\ell,b) = (42\degree,48\degree)$. Our minor axis constraint is inconsistent with estimates from other work. We argue that the inconsistency may arise from each method probing the MW's DM halo at different radial scales, the observational uncertainties, and the LMC's perturbations.
\end{enumerate}
In future work, it is possible to study the effect of orbital phases on our predictions by applying the method to different sections of the stream. With the current star particle resolution of cosmological simulations, the 500 particles required for the method translate to $\sim2\pi$ span in $\theta_r$. Forthcoming suites of cosmological simulations (e.g. Triple Latte FIRE project) have $\sim10$ times better particle resolution, which will not only allow for analyzing the stream by orbital phases but also probing lower-mass streams with stellar masses below $\sim 10^7\Msun$. Moreover, in order to decouple the effects of the observational uncertainties from the LMC's perturbations, a larger volume of data with lower uncertainties will be needed. Future surveys such as SDSS-V \citep{sdss-v}, Rubin Observatory's LSST \citep{lsst}, and the Nancy Grace Roman Space Telescope \cite{wfirst} as well as future data releases from the \textit{Gaia} mission \citep{gaia-mission} will provide unprecedented observational sensitivity which will allow for a sample of streams with full 6-D information that is cleaner, more extensive and has smaller observational uncertainties. This will enable us to better understand the sources of the systematic uncertainties in our method and to apply the method to multiple MW streams that will give us a complete picture of the MW's DM halo.

\section*{Acknowledgements}
NP was supported in part by a Zacheus Daniel fellowship from the University of Pennsylvania.

NP and RES acknowledge support from NASA grant 19-ATP19-0068. The authors thank the Dynamics group at the Center for Computational Astrophysics, Flatiron Institute, for valuable discussion. We also thank Nicolas Garavito-Camargo for providing the data of the LMC-analog from his simulations.

RES additionally acknowledges support from  NSF grant AST-2007232, from the Research Corporation through the Scialog Fellows program on Time Domain Astronomy, and from HST-AR-15809 from the Space Telescope Science Institute (STScI), which is operated by AURA, Inc., under NASA contract NAS5-26555. 

We also would like to thank the Flatiron Institute Scientific Computing Core (SCC) for providing computing resources that made this research possible. Analysis for this paper was carried out on the Flatiron Institute's computing cluster \texttt{rusty}, which is supported by the Simons Foundation.

FIRE simulations were run using XSEDE supported by NSF grant ACI-1548562, Blue Waters via allocation PRAC NSF.1713353 supported by the NSF, and NASA HEC Program through the NAS Division at Ames Research Center.

\software{Astropy (\citealt{astropy:2013}, \citeyear{astropy:2018}, \citeyear{astropy:2022}), IPython \citep{ipython}, Matplotlib \citep{matplotlib}, Numpy \citep{numpy}, Pandas \citep{pandas2}, Scipy \citep{scipy}, \texttt{consistent-trees} \citep{2013ApJ...763...18B}, 
\texttt{rockstar} \citep{2013ApJ...762..109B}, \texttt{halo\_analysis} \citep{2020ascl.soft02014W},
\texttt{gizmo\_analysis} \citep{2020ascl.soft02015W}, NASA's Astrophysics Data System.}

\section*{Data Availability}
The simulated data, including the simulated dwarf galaxy stream, the low-order multipole potential model, and the present-day snapshot, are publicly available as part of the public data release of the FIRE-2 cosmological zoom-in simulations of galaxy formation \citep{2022arXiv220206969W}.

\bibliography{references}

\begin{thebibliography}{}
\expandafter\ifx\csname natexlab\endcsname\relax\def\natexlab#1{#1}\fi
\providecommand{\url}[1]{\href{#1}{#1}}
\providecommand{\dodoi}[1]{doi:~\href{http://doi.org/#1}{\nolinkurl{#1}}}
\providecommand{\doeprint}[1]{\href{http://ascl.net/#1}{\nolinkurl{http://ascl.net/#1}}}
\providecommand{\doarXiv}[1]{\href{https://arxiv.org/abs/#1}{\nolinkurl{https://arxiv.org/abs/#1}}}

\bibitem[{Allgood {et~al.}(2006)Allgood, Flores, Primack, Kravtsov, Wechsler,
  Faltenbacher, \& Bullock}]{allgood2006}
Allgood, B., Flores, R.~A., Primack, J.~R., {et~al.} 2006, Monthly Notices of
  the Royal Astronomical Society, 367, 1781,
  \dodoi{10.1111/j.1365-2966.2006.10094.x}

\bibitem[{Arora {et~al.}(2022)Arora, Sanderson, Panithanpaisal, Cunningham,
  Wetzel, \& Garavito-Camargo}]{arora2022}
Arora, A., Sanderson, R.~E., Panithanpaisal, N., {et~al.} 2022, The
  Astrophysical Journal, 939, 2, \dodoi{10.3847/1538-4357/ac93fb}

\bibitem[{{Astropy Collaboration} {et~al.}(2013){Astropy Collaboration},
  {Robitaille}, {Tollerud}, {Greenfield}, {Droettboom}, {Bray}, {Aldcroft},
  {Davis}, {Ginsburg}, {Price-Whelan}, {Kerzendorf}, {Conley}, {Crighton},
  {Barbary}, {Muna}, {Ferguson}, {Grollier}, {Parikh}, {Nair}, {Unther},
  {Deil}, {Woillez}, {Conseil}, {Kramer}, {Turner}, {Singer}, {Fox}, {Weaver},
  {Zabalza}, {Edwards}, {Azalee Bostroem}, {Burke}, {Casey}, {Crawford},
  {Dencheva}, {Ely}, {Jenness}, {Labrie}, {Lim}, {Pierfederici}, {Pontzen},
  {Ptak}, {Refsdal}, {Servillat}, \& {Streicher}}]{astropy:2013}
{Astropy Collaboration}, {Robitaille}, T.~P., {Tollerud}, E.~J., {et~al.} 2013,
  \aap, 558, A33, \dodoi{10.1051/0004-6361/201322068}

\bibitem[{{Astropy Collaboration} {et~al.}(2018){Astropy Collaboration},
  {Price-Whelan}, {Sip{\H{o}}cz}, {G{\"u}nther}, {Lim}, {Crawford}, {Conseil},
  {Shupe}, {Craig}, {Dencheva}, {Ginsburg}, {Vand erPlas}, {Bradley},
  {P{\'e}rez-Su{\'a}rez}, {de Val-Borro}, {Aldcroft}, {Cruz}, {Robitaille},
  {Tollerud}, {Ardelean}, {Babej}, {Bach}, {Bachetti}, {Bakanov}, {Bamford},
  {Barentsen}, {Barmby}, {Baumbach}, {Berry}, {Biscani}, {Boquien}, {Bostroem},
  {Bouma}, {Brammer}, {Bray}, {Breytenbach}, {Buddelmeijer}, {Burke},
  {Calderone}, {Cano Rodr{\'\i}guez}, {Cara}, {Cardoso}, {Cheedella}, {Copin},
  {Corrales}, {Crichton}, {D'Avella}, {Deil}, {Depagne}, {Dietrich}, {Donath},
  {Droettboom}, {Earl}, {Erben}, {Fabbro}, {Ferreira}, {Finethy}, {Fox},
  {Garrison}, {Gibbons}, {Goldstein}, {Gommers}, {Greco}, {Greenfield},
  {Groener}, {Grollier}, {Hagen}, {Hirst}, {Homeier}, {Horton}, {Hosseinzadeh},
  {Hu}, {Hunkeler}, {Ivezi{\'c}}, {Jain}, {Jenness}, {Kanarek}, {Kendrew},
  {Kern}, {Kerzendorf}, {Khvalko}, {King}, {Kirkby}, {Kulkarni}, {Kumar},
  {Lee}, {Lenz}, {Littlefair}, {Ma}, {Macleod}, {Mastropietro}, {McCully},
  {Montagnac}, {Morris}, {Mueller}, {Mumford}, {Muna}, {Murphy}, {Nelson},
  {Nguyen}, {Ninan}, {N{\"o}the}, {Ogaz}, {Oh}, {Parejko}, {Parley}, {Pascual},
  {Patil}, {Patil}, {Plunkett}, {Prochaska}, {Rastogi}, {Reddy Janga},
  {Sabater}, {Sakurikar}, {Seifert}, {Sherbert}, {Sherwood-Taylor}, {Shih},
  {Sick}, {Silbiger}, {Singanamalla}, {Singer}, {Sladen}, {Sooley},
  {Sornarajah}, {Streicher}, {Teuben}, {Thomas}, {Tremblay}, {Turner},
  {Terr{\'o}n}, {van Kerkwijk}, {de la Vega}, {Watkins}, {Weaver}, {Whitmore},
  {Woillez}, {Zabalza}, \& {Astropy Contributors}}]{astropy:2018}
{Astropy Collaboration}, {Price-Whelan}, A.~M., {Sip{\H{o}}cz}, B.~M., {et~al.}
  2018, \aj, 156, 123, \dodoi{10.3847/1538-3881/aabc4f}

\bibitem[{{Astropy Collaboration} {et~al.}(2022){Astropy Collaboration},
  {Price-Whelan}, {Lim}, {Earl}, {Starkman}, {Bradley}, {Shupe}, {Patil},
  {Corrales}, {Brasseur}, {N{"o}the}, {Donath}, {Tollerud}, {Morris},
  {Ginsburg}, {Vaher}, {Weaver}, {Tocknell}, {Jamieson}, {van Kerkwijk},
  {Robitaille}, {Merry}, {Bachetti}, {G{"u}nther}, {Aldcroft},
  {Alvarado-Montes}, {Archibald}, {B{'o}di}, {Bapat}, {Barentsen}, {Baz{'a}n},
  {Biswas}, {Boquien}, {Burke}, {Cara}, {Cara}, {Conroy}, {Conseil}, {Craig},
  {Cross}, {Cruz}, {D'Eugenio}, {Dencheva}, {Devillepoix}, {Dietrich},
  {Eigenbrot}, {Erben}, {Ferreira}, {Foreman-Mackey}, {Fox}, {Freij}, {Garg},
  {Geda}, {Glattly}, {Gondhalekar}, {Gordon}, {Grant}, {Greenfield}, {Groener},
  {Guest}, {Gurovich}, {Handberg}, {Hart}, {Hatfield-Dodds}, {Homeier},
  {Hosseinzadeh}, {Jenness}, {Jones}, {Joseph}, {Kalmbach}, {Karamehmetoglu},
  {Ka{l}uszy{'n}ski}, {Kelley}, {Kern}, {Kerzendorf}, {Koch}, {Kulumani},
  {Lee}, {Ly}, {Ma}, {MacBride}, {Maljaars}, {Muna}, {Murphy}, {Norman},
  {O'Steen}, {Oman}, {Pacifici}, {Pascual}, {Pascual-Granado}, {Patil},
  {Perren}, {Pickering}, {Rastogi}, {Roulston}, {Ryan}, {Rykoff}, {Sabater},
  {Sakurikar}, {Salgado}, {Sanghi}, {Saunders}, {Savchenko}, {Schwardt},
  {Seifert-Eckert}, {Shih}, {Jain}, {Shukla}, {Sick}, {Simpson},
  {Singanamalla}, {Singer}, {Singhal}, {Sinha}, {Sip{H{o}}cz}, {Spitler},
  {Stansby}, {Streicher}, {{{S}}umak}, {Swinbank}, {Taranu}, {Tewary},
  {Tremblay}, {Val-Borro}, {Van Kooten}, {Vasovi{'c}}, {Verma}, {de Miranda
  Cardoso}, {Williams}, {Wilson}, {Winkel}, {Wood-Vasey}, {Xue}, {Yoachim},
  {Zhang}, {Zonca}, \& {Astropy Project Contributors}}]{astropy:2022}
{Astropy Collaboration}, {Price-Whelan}, A.~M., {Lim}, P.~L., {et~al.} 2022,
  apj, 935, 167, \dodoi{10.3847/1538-4357/ac7c74}

\bibitem[{{Behroozi} {et~al.}(2013{\natexlab{a}}){Behroozi}, {Wechsler}, \&
  {Wu}}]{2013ApJ...762..109B}
{Behroozi}, P.~S., {Wechsler}, R.~H., \& {Wu}, H.-Y. 2013{\natexlab{a}}, \apj,
  762, 109, \dodoi{10.1088/0004-637X/762/2/109}

\bibitem[{{Behroozi} {et~al.}(2013{\natexlab{b}}){Behroozi}, {Wechsler}, {Wu},
  {Busha}, {Klypin}, \& {Primack}}]{2013ApJ...763...18B}
{Behroozi}, P.~S., {Wechsler}, R.~H., {Wu}, H.-Y., {et~al.} 2013{\natexlab{b}},
  \apj, 763, 18, \dodoi{10.1088/0004-637X/763/1/18}

\bibitem[{{Belokurov} {et~al.}(2018){Belokurov}, {Erkal}, {Evans}, {Koposov},
  \& {Deason}}]{Belokurov2018}
{Belokurov}, V., {Erkal}, D., {Evans}, N.~W., {Koposov}, S.~E., \& {Deason},
  A.~J. 2018, \mnras, 478, 611, \dodoi{10.1093/mnras/sty982}

\bibitem[{{Besla} {et~al.}(2007){Besla}, {Kallivayalil}, {Hernquist},
  {Robertson}, {Cox}, {van der Marel}, \& {Alcock}}]{besla2007}
{Besla}, G., {Kallivayalil}, N., {Hernquist}, L., {et~al.} 2007, \apj, 668,
  949, \dodoi{10.1086/521385}

\bibitem[{{Binney}(2012)}]{2012MNRAS.426.1324B}
{Binney}, J. 2012, \mnras, 426, 1324, \dodoi{10.1111/j.1365-2966.2012.21757.x}

\bibitem[{{Bonaca} {et~al.}(2019){Bonaca}, {Hogg}, {Price-Whelan}, \&
  {Conroy}}]{bonanca2019}
{Bonaca}, A., {Hogg}, D.~W., {Price-Whelan}, A.~M., \& {Conroy}, C. 2019, \apj,
  880, 38, \dodoi{10.3847/1538-4357/ab2873}

\bibitem[{{Bovy}(2015)}]{bovy2015}
{Bovy}, J. 2015, \apjs, 216, 29, \dodoi{10.1088/0067-0049/216/2/29}

\bibitem[{{Buist} \& {Helmi}(2017)}]{buist2017}
{Buist}, H. J.~T., \& {Helmi}, A. 2017, \aap, 601, A37,
  \dodoi{10.1051/0004-6361/201527930}

\bibitem[{{Carlberg} \& {Grillmair}(2013)}]{carlberg2013}
{Carlberg}, R.~G., \& {Grillmair}, C.~J. 2013, \apj, 768, 171,
  \dodoi{10.1088/0004-637X/768/2/171}

\bibitem[{{Clark} {et~al.}(2020){Clark}, {Peek}, {Putman}, {Schudel}, \&
  {Jaspers}}]{2020ascl.soft03005C}
{Clark}, S.~E., {Peek}, J., {Putman}, M., {Schudel}, L., \& {Jaspers}, R. 2020,
  {RHT: Rolling Hough Transform}, Astrophysics Source Code Library, record
  ascl:2003.005.
\newblock \doeprint{2003.005}

\bibitem[{{Conroy} {et~al.}(2021){Conroy}, {Naidu}, {Garavito-Camargo},
  {Besla}, {Zaritsky}, {Bonaca}, \& {Johnson}}]{conroy2021}
{Conroy}, C., {Naidu}, R.~P., {Garavito-Camargo}, N., {et~al.} 2021, \nat, 592,
  534, \dodoi{10.1038/s41586-021-03385-7}

\bibitem[{{Cropper} {et~al.}(2018){Cropper}, {Katz}, {Sartoretti}, {Prusti},
  {de Bruijne}, {Chassat}, {Charvet}, {Boyadjian}, {Perryman}, {Sarri}, {Gare},
  {Erdmann}, {Munari}, {Zwitter}, {Wilkinson}, {Arenou}, {Vallenari},
  {G{\'o}mez}, {Panuzzo}, {Seabroke}, {Allende Prieto}, {Benson}, {Marchal},
  {Huckle}, {Smith}, {Dolding}, {Jan{\ss}en}, {Viala}, {Blomme}, {Baker},
  {Boudreault}, {Crifo}, {Soubiran}, {Fr{\'e}mat}, {Jasniewicz}, {Guerrier},
  {Guy}, {Turon}, {Jean-Antoine-Piccolo}, {Th{\'e}venin}, {David}, {Gosset}, \&
  {Damerdji}}]{gaia-rvs}
{Cropper}, M., {Katz}, D., {Sartoretti}, P., {et~al.} 2018, \aap, 616, A5,
  \dodoi{10.1051/0004-6361/201832763}

\bibitem[{{Cui} {et~al.}(2012){Cui}, {Zhao}, {Chu}, {Li}, {Li}, {Zhang}, {Su},
  {Yao}, {Wang}, {Xing}, {Li}, {Zhu}, {Wang}, {Gu}, {Luo}, {Xu}, {Zhang},
  {Liu}, {Zhang}, {Yang}, {Cao}, {Chen}, {Chen}, {Chen}, {Chen}, {Chu}, {Feng},
  {Gong}, {Hou}, {Hu}, {Hu}, {Hu}, {Jia}, {Jiang}, {Jiang}, {Jiang}, {Jin},
  {Li}, {Li}, {Li}, {Liu}, {Liu}, {Lu}, {Mao}, {Men}, {Qi}, {Qi}, {Shi},
  {Tang}, {Tao}, {Wang}, {Wang}, {Wang}, {Wang}, {Wang}, {Wang}, {Wang},
  {Wang}, {Wang}, {Wang}, {Wang}, {Wang}, {Xu}, {Xu}, {Yang}, {Yu}, {Yuan},
  {Yuan}, {Zhai}, {Zhang}, {Zhang}, {Zhang}, {Zhao}, {Zhou}, {Zhou}, {Zhu}, \&
  {Zou}}]{lamost}
{Cui}, X.-Q., {Zhao}, Y.-H., {Chu}, Y.-Q., {et~al.} 2012, Research in Astronomy
  and Astrophysics, 12, 1197, \dodoi{10.1088/1674-4527/12/9/003}

\bibitem[{{Cunningham} {et~al.}(2020){Cunningham}, {Garavito-Camargo},
  {Deason}, {Johnston}, {Erkal}, {Laporte}, {Besla}, {Luger}, \&
  {Sanderson}}]{cunningham2020}
{Cunningham}, E.~C., {Garavito-Camargo}, N., {Deason}, A.~J., {et~al.} 2020,
  \apj, 898, 4, \dodoi{10.3847/1538-4357/ab9b88}

\bibitem[{{Debattista} {et~al.}(2019){Debattista}, {Gonzalez}, {Sanderson},
  {El-Badry}, {Garrison-Kimmel}, {Wetzel}, {Faucher-Gigu{\`e}re}, \&
  {Hopkins}}]{2019MNRAS.485.5073D}
{Debattista}, V.~P., {Gonzalez}, O.~A., {Sanderson}, R.~E., {et~al.} 2019,
  \mnras, 485, 5073, \dodoi{10.1093/mnras/stz746}

\bibitem[{{Debattista} {et~al.}(2013){Debattista}, {Ro{\v{s}}kar}, {Valluri},
  {Quinn}, {Moore}, \& {Wadsley}}]{debattista2013}
{Debattista}, V.~P., {Ro{\v{s}}kar}, R., {Valluri}, M., {et~al.} 2013, \mnras,
  434, 2971, \dodoi{10.1093/mnras/stt1217}

\bibitem[{Duda \& Hart(1972)}]{10.1145/361237.361242}
Duda, R.~O., \& Hart, P.~E. 1972, Commun. ACM, 15, 11–15,
  \dodoi{10.1145/361237.361242}

\bibitem[{{Emami} {et~al.}(2021){Emami}, {Genel}, {Hernquist}, {Alcock},
  {Bose}, {Weinberger}, {Vogelsberger}, {Marinacci}, {Loeb}, {Torrey}, \&
  {Forbes}}]{emami2021}
{Emami}, R., {Genel}, S., {Hernquist}, L., {et~al.} 2021, \apj, 913, 36,
  \dodoi{10.3847/1538-4357/abf147}

\bibitem[{{Erkal} {et~al.}(2019){Erkal}, {Belokurov}, {Laporte}, {Koposov},
  {Li}, {Grillmair}, {Kallivayalil}, {Price-Whelan}, {Evans}, {Hawkins},
  {Hendel}, {Mateu}, {Navarro}, {del Pino}, {Slater}, {Sohn}, \& {Orphan Aspen
  Treasury Collaboration}}]{erkal2019}
{Erkal}, D., {Belokurov}, V., {Laporte}, C.~F.~P., {et~al.} 2019, \mnras, 487,
  2685, \dodoi{10.1093/mnras/stz1371}

\bibitem[{{Gaia Collaboration} {et~al.}(2016){Gaia Collaboration}, {Prusti},
  {de Bruijne}, {Brown}, {Vallenari}, {Babusiaux}, {Bailer-Jones}, {Bastian},
  {Biermann}, {Evans}, {Eyer}, {Jansen}, {Jordi}, {Klioner}, {Lammers},
  {Lindegren}, {Luri}, {Mignard}, {Milligan}, {Panem}, {Poinsignon},
  {Pourbaix}, {Randich}, {Sarri}, {Sartoretti}, {Siddiqui}, {Soubiran},
  {Valette}, {van Leeuwen}, {Walton}, {Aerts}, {Arenou}, {Cropper}, {Drimmel},
  {H{\o}g}, {Katz}, {Lattanzi}, {O'Mullane}, {Grebel}, {Holland}, {Huc},
  {Passot}, {Bramante}, {Cacciari}, {Casta{\~n}eda}, {Chaoul}, {Cheek}, {De
  Angeli}, {Fabricius}, {Guerra}, {Hern{\'a}ndez}, {Jean-Antoine-Piccolo},
  {Masana}, {Messineo}, {Mowlavi}, {Nienartowicz}, {Ord{\'o}{\~n}ez-Blanco},
  {Panuzzo}, {Portell}, {Richards}, {Riello}, {Seabroke}, {Tanga},
  {Th{\'e}venin}, {Torra}, {Els}, {Gracia-Abril}, {Comoretto},
  {Garcia-Reinaldos}, {Lock}, {Mercier}, {Altmann}, {Andrae}, {Astraatmadja},
  {Bellas-Velidis}, {Benson}, {Berthier}, {Blomme}, {Busso}, {Carry},
  {Cellino}, {Clementini}, {Cowell}, {Creevey}, {Cuypers}, {Davidson}, {De
  Ridder}, {de Torres}, {Delchambre}, {Dell'Oro}, {Ducourant}, {Fr{\'e}mat},
  {Garc{\'\i}a-Torres}, {Gosset}, {Halbwachs}, {Hambly}, {Harrison}, {Hauser},
  {Hestroffer}, {Hodgkin}, {Huckle}, {Hutton}, {Jasniewicz}, {Jordan},
  {Kontizas}, {Korn}, {Lanzafame}, {Manteiga}, {Moitinho}, {Muinonen},
  {Osinde}, {Pancino}, {Pauwels}, {Petit}, {Recio-Blanco}, {Robin}, {Sarro},
  {Siopis}, {Smith}, {Smith}, {Sozzetti}, {Thuillot}, {van Reeven}, {Viala},
  {Abbas}, {Abreu Aramburu}, {Accart}, {Aguado}, {Allan}, {Allasia},
  {Altavilla}, {{\'A}lvarez}, {Alves}, {Anderson}, {Andrei}, {Anglada Varela},
  {Antiche}, {Antoja}, {Ant{\'o}n}, {Arcay}, {Atzei}, {Ayache}, {Bach},
  {Baker}, {Balaguer-N{\'u}{\~n}ez}, {Barache}, {Barata}, {Barbier}, {Barblan},
  {Baroni}, {Barrado y Navascu{\'e}s}, {Barros}, {Barstow}, {Becciani},
  {Bellazzini}, {Bellei}, {Bello Garc{\'\i}a}, {Belokurov}, {Bendjoya},
  {Berihuete}, {Bianchi}, {Bienaym{\'e}}, {Billebaud}, {Blagorodnova},
  {Blanco-Cuaresma}, {Boch}, {Bombrun}, {Borrachero}, {Bouquillon}, {Bourda},
  {Bouy}, {Bragaglia}, {Breddels}, {Brouillet}, {Br{\"u}semeister},
  {Bucciarelli}, {Budnik}, {Burgess}, {Burgon}, {Burlacu}, {Busonero}, {Buzzi},
  {Caffau}, {Cambras}, {Campbell}, {Cancelliere}, {Cantat-Gaudin}, {Carlucci},
  {Carrasco}, {Castellani}, {Charlot}, {Charnas}, {Charvet}, {Chassat},
  {Chiavassa}, {Clotet}, {Cocozza}, {Collins}, {Collins}, {Costigan}, {Crifo},
  {Cross}, {Crosta}, {Crowley}, {Dafonte}, {Damerdji}, {Dapergolas}, {David},
  {David}, {De Cat}, {de Felice}, {de Laverny}, {De Luise}, {De March}, {de
  Martino}, {de Souza}, {Debosscher}, {del Pozo}, {Delbo}, {Delgado},
  {Delgado}, {di Marco}, {Di Matteo}, {Diakite}, {Distefano}, {Dolding}, {Dos
  Anjos}, {Drazinos}, {Dur{\'a}n}, {Dzigan}, {Ecale}, {Edvardsson}, {Enke},
  {Erdmann}, {Escolar}, {Espina}, {Evans}, {Eynard Bontemps}, {Fabre},
  {Fabrizio}, {Faigler}, {Falc{\~a}o}, {Farr{\`a}s Casas}, {Faye}, {Federici},
  {Fedorets}, {Fern{\'a}ndez-Hern{\'a}ndez}, {Fernique}, {Fienga}, {Figueras},
  {Filippi}, {Findeisen}, {Fonti}, {Fouesneau}, {Fraile}, {Fraser}, {Fuchs},
  {Furnell}, {Gai}, {Galleti}, {Galluccio}, {Garabato}, {Garc{\'\i}a-Sedano},
  {Gar{\'e}}, {Garofalo}, {Garralda}, {Gavras}, {Gerssen}, {Geyer}, {Gilmore},
  {Girona}, {Giuffrida}, {Gomes}, {Gonz{\'a}lez-Marcos},
  {Gonz{\'a}lez-N{\'u}{\~n}ez}, {Gonz{\'a}lez-Vidal}, {Granvik}, {Guerrier},
  {Guillout}, {Guiraud}, {G{\'u}rpide}, {Guti{\'e}rrez-S{\'a}nchez}, {Guy},
  {Haigron}, {Hatzidimitriou}, {Haywood}, {Heiter}, {Helmi}, {Hobbs},
  {Hofmann}, {Holl}, {Holland}, {Hunt}, {Hypki}, {Icardi}, {Irwin}, {Jevardat
  de Fombelle}, {Jofr{\'e}}, {Jonker}, {Jorissen}, {Julbe}, {Karampelas},
  {Kochoska}, {Kohley}, {Kolenberg}, {Kontizas}, {Koposov}, {Kordopatis},
  {Koubsky}, {Kowalczyk}, {Krone-Martins}, {Kudryashova}, {Kull}, {Bachchan},
  {Lacoste-Seris}, {Lanza}, {Lavigne}, {Le Poncin-Lafitte}, {Lebreton},
  {Lebzelter}, {Leccia}, {Leclerc}, {Lecoeur-Taibi}, {Lemaitre}, {Lenhardt},
  {Leroux}, {Liao}, {Licata}, {Lindstr{\o}m}, {Lister}, {Livanou}, {Lobel},
  {L{\"o}ffler}, {L{\'o}pez}, {Lopez-Lozano}, {Lorenz}, {Loureiro},
  {MacDonald}, {Magalh{\~a}es Fernandes}, {Managau}, {Mann}, {Mantelet},
  {Marchal}, {Marchant}, {Marconi}, {Marie}, {Marinoni}, {Marrese},
  {Marschalk{\'o}}, {Marshall}, {Mart{\'\i}n-Fleitas}, {Martino}, {Mary},
  {Matijevi{\v{c}}}, {Mazeh}, {McMillan}, {Messina}, {Mestre}, {Michalik},
  {Millar}, {Miranda}, {Molina}, {Molinaro}, {Molinaro}, {Moln{\'a}r},
  {Moniez}, {Montegriffo}, {Monteiro}, {Mor}, {Mora}, {Morbidelli}, {Morel},
  {Morgenthaler}, {Morley}, {Morris}, {Mulone}, {Muraveva}, {Musella},
  {Narbonne}, {Nelemans}, {Nicastro}, {Noval}, {Ord{\'e}novic},
  {Ordieres-Mer{\'e}}, {Osborne}, {Pagani}, {Pagano}, {Pailler}, {Palacin},
  {Palaversa}, {Parsons}, {Paulsen}, {Pecoraro}, {Pedrosa}, {Pentik{\"a}inen},
  {Pereira}, {Pichon}, {Piersimoni}, {Pineau}, {Plachy}, {Plum}, {Poujoulet},
  {Pr{\v{s}}a}, {Pulone}, {Ragaini}, {Rago}, {Rambaux}, {Ramos-Lerate},
  {Ranalli}, {Rauw}, {Read}, {Regibo}, {Renk}, {Reyl{\'e}}, {Ribeiro},
  {Rimoldini}, {Ripepi}, {Riva}, {Rixon}, {Roelens}, {Romero-G{\'o}mez},
  {Rowell}, {Royer}, {Rudolph}, {Ruiz-Dern}, {Sadowski}, {Sagrist{\`a}
  Sell{\'e}s}, {Sahlmann}, {Salgado}, {Salguero}, {Sarasso}, {Savietto},
  {Schnorhk}, {Schultheis}, {Sciacca}, {Segol}, {Segovia}, {Segransan},
  {Serpell}, {Shih}, {Smareglia}, {Smart}, {Smith}, {Solano}, {Solitro},
  {Sordo}, {Soria Nieto}, {Souchay}, {Spagna}, {Spoto}, {Stampa}, {Steele},
  {Steidelm{\"u}ller}, {Stephenson}, {Stoev}, {Suess}, {S{\"u}veges}, {Surdej},
  {Szabados}, {Szegedi-Elek}, {Tapiador}, {Taris}, {Tauran}, {Taylor},
  {Teixeira}, {Terrett}, {Tingley}, {Trager}, {Turon}, {Ulla}, {Utrilla},
  {Valentini}, {van Elteren}, {Van Hemelryck}, {van Leeuwen}, {Varadi},
  {Vecchiato}, {Veljanoski}, {Via}, {Vicente}, {Vogt}, {Voss}, {Votruba},
  {Voutsinas}, {Walmsley}, {Weiler}, {Weingrill}, {Werner}, {Wevers},
  {Whitehead}, {Wyrzykowski}, {Yoldas}, {{\v{Z}}erjal}, {Zucker}, {Zurbach},
  {Zwitter}, {Alecu}, {Allen}, {Allende Prieto}, {Amorim},
  {Anglada-Escud{\'e}}, {Arsenijevic}, {Azaz}, {Balm}, {Beck}, {Bernstein},
  {Bigot}, {Bijaoui}, {Blasco}, {Bonfigli}, {Bono}, {Boudreault}, {Bressan},
  {Brown}, {Brunet}, {Bunclark}, {Buonanno}, {Butkevich}, {Carret}, {Carrion},
  {Chemin}, {Ch{\'e}reau}, {Corcione}, {Darmigny}, {de Boer}, {de Teodoro}, {de
  Zeeuw}, {Delle Luche}, {Domingues}, {Dubath}, {Fodor}, {Fr{\'e}zouls},
  {Fries}, {Fustes}, {Fyfe}, {Gallardo}, {Gallegos}, {Gardiol}, {Gebran},
  {Gomboc}, {G{\'o}mez}, {Grux}, {Gueguen}, {Heyrovsky}, {Hoar}, {Iannicola},
  {Isasi Parache}, {Janotto}, {Joliet}, {Jonckheere}, {Keil}, {Kim},
  {Klagyivik}, {Klar}, {Knude}, {Kochukhov}, {Kolka}, {Kos}, {Kutka}, {Lainey},
  {LeBouquin}, {Liu}, {Loreggia}, {Makarov}, {Marseille}, {Martayan},
  {Martinez-Rubi}, {Massart}, {Meynadier}, {Mignot}, {Munari}, {Nguyen},
  {Nordlander}, {Ocvirk}, {O'Flaherty}, {Olias Sanz}, {Ortiz}, {Osorio},
  {Oszkiewicz}, {Ouzounis}, {Palmer}, {Park}, {Pasquato}, {Peltzer}, {Peralta},
  {P{\'e}turaud}, {Pieniluoma}, {Pigozzi}, {Poels}, {Prat}, {Prod'homme},
  {Raison}, {Rebordao}, {Risquez}, {Rocca-Volmerange}, {Rosen}, {Ruiz-Fuertes},
  {Russo}, {Sembay}, {Serraller Vizcaino}, {Short}, {Siebert}, {Silva},
  {Sinachopoulos}, {Slezak}, {Soffel}, {Sosnowska}, {Strai{\v{z}}ys}, {ter
  Linden}, {Terrell}, {Theil}, {Tiede}, {Troisi}, {Tsalmantza}, {Tur},
  {Vaccari}, {Vachier}, {Valles}, {Van Hamme}, {Veltz}, {Virtanen}, {Wallut},
  {Wichmann}, {Wilkinson}, {Ziaeepour}, \& {Zschocke}}]{gaia-mission}
{Gaia Collaboration}, {Prusti}, T., {de Bruijne}, J.~H.~J., {et~al.} 2016,
  \aap, 595, A1, \dodoi{10.1051/0004-6361/201629272}

\bibitem[{{Gaia Collaboration} {et~al.}(2018){Gaia Collaboration}, {Brown},
  {Vallenari}, {Prusti}, {de Bruijne}, {Babusiaux}, {Bailer-Jones}, {Biermann},
  {Evans}, {Eyer}, {Jansen}, {Jordi}, {Klioner}, {Lammers}, {Lindegren},
  {Luri}, {Mignard}, {Panem}, {Pourbaix}, {Randich}, {Sartoretti}, {Siddiqui},
  {Soubiran}, {van Leeuwen}, {Walton}, {Arenou}, {Bastian}, {Cropper},
  {Drimmel}, {Katz}, {Lattanzi}, {Bakker}, {Cacciari}, {Casta{\~n}eda},
  {Chaoul}, {Cheek}, {De Angeli}, {Fabricius}, {Guerra}, {Holl}, {Masana},
  {Messineo}, {Mowlavi}, {Nienartowicz}, {Panuzzo}, {Portell}, {Riello},
  {Seabroke}, {Tanga}, {Th{\'e}venin}, {Gracia-Abril}, {Comoretto},
  {Garcia-Reinaldos}, {Teyssier}, {Altmann}, {Andrae}, {Audard},
  {Bellas-Velidis}, {Benson}, {Berthier}, {Blomme}, {Burgess}, {Busso},
  {Carry}, {Cellino}, {Clementini}, {Clotet}, {Creevey}, {Davidson}, {De
  Ridder}, {Delchambre}, {Dell'Oro}, {Ducourant},
  {Fern{\'a}ndez-Hern{\'a}ndez}, {Fouesneau}, {Fr{\'e}mat}, {Galluccio},
  {Garc{\'\i}a-Torres}, {Gonz{\'a}lez-N{\'u}{\~n}ez}, {Gonz{\'a}lez-Vidal},
  {Gosset}, {Guy}, {Halbwachs}, {Hambly}, {Harrison}, {Hern{\'a}ndez},
  {Hestroffer}, {Hodgkin}, {Hutton}, {Jasniewicz}, {Jean-Antoine-Piccolo},
  {Jordan}, {Korn}, {Krone-Martins}, {Lanzafame}, {Lebzelter}, {L{\"o}ffler},
  {Manteiga}, {Marrese}, {Mart{\'\i}n-Fleitas}, {Moitinho}, {Mora}, {Muinonen},
  {Osinde}, {Pancino}, {Pauwels}, {Petit}, {Recio-Blanco}, {Richards},
  {Rimoldini}, {Robin}, {Sarro}, {Siopis}, {Smith}, {Sozzetti}, {S{\"u}veges},
  {Torra}, {van Reeven}, {Abbas}, {Abreu Aramburu}, {Accart}, {Aerts},
  {Altavilla}, {{\'A}lvarez}, {Alvarez}, {Alves}, {Anderson}, {Andrei},
  {Anglada Varela}, {Antiche}, {Antoja}, {Arcay}, {Astraatmadja}, {Bach},
  {Baker}, {Balaguer-N{\'u}{\~n}ez}, {Balm}, {Barache}, {Barata}, {Barbato},
  {Barblan}, {Barklem}, {Barrado}, {Barros}, {Barstow}, {Bartholom{\'e}
  Mu{\~n}oz}, {Bassilana}, {Becciani}, {Bellazzini}, {Berihuete}, {Bertone},
  {Bianchi}, {Bienaym{\'e}}, {Blanco-Cuaresma}, {Boch}, {Boeche}, {Bombrun},
  {Borrachero}, {Bossini}, {Bouquillon}, {Bourda}, {Bragaglia}, {Bramante},
  {Breddels}, {Bressan}, {Brouillet}, {Br{\"u}semeister}, {Brugaletta},
  {Bucciarelli}, {Burlacu}, {Busonero}, {Butkevich}, {Buzzi}, {Caffau},
  {Cancelliere}, {Cannizzaro}, {Cantat-Gaudin}, {Carballo}, {Carlucci},
  {Carrasco}, {Casamiquela}, {Castellani}, {Castro-Ginard}, {Charlot},
  {Chemin}, {Chiavassa}, {Cocozza}, {Costigan}, {Cowell}, {Crifo}, {Crosta},
  {Crowley}, {Cuypers}, {Dafonte}, {Damerdji}, {Dapergolas}, {David}, {David},
  {de Laverny}, {De Luise}, {De March}, {de Martino}, {de Souza}, {de Torres},
  {Debosscher}, {del Pozo}, {Delbo}, {Delgado}, {Delgado}, {Di Matteo},
  {Diakite}, {Diener}, {Distefano}, {Dolding}, {Drazinos}, {Dur{\'a}n},
  {Edvardsson}, {Enke}, {Eriksson}, {Esquej}, {Eynard Bontemps}, {Fabre},
  {Fabrizio}, {Faigler}, {Falc{\~a}o}, {Farr{\`a}s Casas}, {Federici},
  {Fedorets}, {Fernique}, {Figueras}, {Filippi}, {Findeisen}, {Fonti},
  {Fraile}, {Fraser}, {Fr{\'e}zouls}, {Gai}, {Galleti}, {Garabato},
  {Garc{\'\i}a-Sedano}, {Garofalo}, {Garralda}, {Gavel}, {Gavras}, {Gerssen},
  {Geyer}, {Giacobbe}, {Gilmore}, {Girona}, {Giuffrida}, {Glass}, {Gomes},
  {Granvik}, {Gueguen}, {Guerrier}, {Guiraud}, {Guti{\'e}rrez-S{\'a}nchez},
  {Haigron}, {Hatzidimitriou}, {Hauser}, {Haywood}, {Heiter}, {Helmi}, {Heu},
  {Hilger}, {Hobbs}, {Hofmann}, {Holland}, {Huckle}, {Hypki}, {Icardi},
  {Jan{\ss}en}, {Jevardat de Fombelle}, {Jonker}, {Juh{\'a}sz}, {Julbe},
  {Karampelas}, {Kewley}, {Klar}, {Kochoska}, {Kohley}, {Kolenberg},
  {Kontizas}, {Kontizas}, {Koposov}, {Kordopatis}, {Kostrzewa-Rutkowska},
  {Koubsky}, {Lambert}, {Lanza}, {Lasne}, {Lavigne}, {Le Fustec}, {Le
  Poncin-Lafitte}, {Lebreton}, {Leccia}, {Leclerc}, {Lecoeur-Taibi},
  {Lenhardt}, {Leroux}, {Liao}, {Licata}, {Lindstr{\o}m}, {Lister}, {Livanou},
  {Lobel}, {L{\'o}pez}, {Managau}, {Mann}, {Mantelet}, {Marchal}, {Marchant},
  {Marconi}, {Marinoni}, {Marschalk{\'o}}, {Marshall}, {Martino}, {Marton},
  {Mary}, {Massari}, {Matijevi{\v{c}}}, {Mazeh}, {McMillan}, {Messina},
  {Michalik}, {Millar}, {Molina}, {Molinaro}, {Moln{\'a}r}, {Montegriffo},
  {Mor}, {Morbidelli}, {Morel}, {Morris}, {Mulone}, {Muraveva}, {Musella},
  {Nelemans}, {Nicastro}, {Noval}, {O'Mullane}, {Ord{\'e}novic},
  {Ord{\'o}{\~n}ez-Blanco}, {Osborne}, {Pagani}, {Pagano}, {Pailler},
  {Palacin}, {Palaversa}, {Panahi}, {Pawlak}, {Piersimoni}, {Pineau}, {Plachy},
  {Plum}, {Poggio}, {Poujoulet}, {Pr{\v{s}}a}, {Pulone}, {Racero}, {Ragaini},
  {Rambaux}, {Ramos-Lerate}, {Regibo}, {Reyl{\'e}}, {Riclet}, {Ripepi}, {Riva},
  {Rivard}, {Rixon}, {Roegiers}, {Roelens}, {Romero-G{\'o}mez}, {Rowell},
  {Royer}, {Ruiz-Dern}, {Sadowski}, {Sagrist{\`a} Sell{\'e}s}, {Sahlmann},
  {Salgado}, {Salguero}, {Sanna}, {Santana-Ros}, {Sarasso}, {Savietto},
  {Schultheis}, {Sciacca}, {Segol}, {Segovia}, {S{\'e}gransan}, {Shih},
  {Siltala}, {Silva}, {Smart}, {Smith}, {Solano}, {Solitro}, {Sordo}, {Soria
  Nieto}, {Souchay}, {Spagna}, {Spoto}, {Stampa}, {Steele},
  {Steidelm{\"u}ller}, {Stephenson}, {Stoev}, {Suess}, {Surdej}, {Szabados},
  {Szegedi-Elek}, {Tapiador}, {Taris}, {Tauran}, {Taylor}, {Teixeira},
  {Terrett}, {Teyssandier}, {Thuillot}, {Titarenko}, {Torra Clotet}, {Turon},
  {Ulla}, {Utrilla}, {Uzzi}, {Vaillant}, {Valentini}, {Valette}, {van Elteren},
  {Van Hemelryck}, {van Leeuwen}, {Vaschetto}, {Vecchiato}, {Veljanoski},
  {Viala}, {Vicente}, {Vogt}, {von Essen}, {Voss}, {Votruba}, {Voutsinas},
  {Walmsley}, {Weiler}, {Wertz}, {Wevers}, {Wyrzykowski}, {Yoldas},
  {{\v{Z}}erjal}, {Ziaeepour}, {Zorec}, {Zschocke}, {Zucker}, {Zurbach}, \&
  {Zwitter}}]{gaia-dr2}
{Gaia Collaboration}, {Brown}, A.~G.~A., {Vallenari}, A., {et~al.} 2018, \aap,
  616, A1, \dodoi{10.1051/0004-6361/201833051}

\bibitem[{{Garavito-Camargo} {et~al.}(2019){Garavito-Camargo}, {Besla},
  {Laporte}, {Johnston}, {G{\'o}mez}, \& {Watkins}}]{garavito2019}
{Garavito-Camargo}, N., {Besla}, G., {Laporte}, C. F.~P., {et~al.} 2019, \apj,
  884, 51, \dodoi{10.3847/1538-4357/ab32eb}

\bibitem[{{Garavito-Camargo} {et~al.}(2021){Garavito-Camargo}, {Besla},
  {Laporte}, {Price-Whelan}, {Cunningham}, {Johnston}, {Weinberg}, \&
  {G{\'o}mez}}]{garavito2021}
---. 2021, \apj, 919, 109, \dodoi{10.3847/1538-4357/ac0b44}

\bibitem[{{Garrison-Kimmel} {et~al.}(2019){Garrison-Kimmel}, {Hopkins},
  {Wetzel}, {Bullock}, {Boylan-Kolchin}, {Kere{\v{s}}}, {Faucher-Gigu{\`e}re},
  {El-Badry}, {Lamberts}, {Quataert}, \& {Sand erson}}]{Garrison-Kimmel2019}
{Garrison-Kimmel}, S., {Hopkins}, P.~F., {Wetzel}, A., {et~al.} 2019, \mnras,
  487, 1380, \dodoi{10.1093/mnras/stz1317}

\bibitem[{{Gibbons} {et~al.}(2017){Gibbons}, {Belokurov}, \&
  {Evans}}]{gibbons2017}
{Gibbons}, S.~L.~J., {Belokurov}, V., \& {Evans}, N.~W. 2017, \mnras, 464, 794,
  \dodoi{10.1093/mnras/stw2328}

\bibitem[{{G{\'o}rski} {et~al.}(2005){G{\'o}rski}, {Hivon}, {Banday},
  {Wandelt}, {Hansen}, {Reinecke}, \& {Bartelmann}}]{2005ApJ...622..759G}
{G{\'o}rski}, K.~M., {Hivon}, E., {Banday}, A.~J., {et~al.} 2005, \apj, 622,
  759, \dodoi{10.1086/427976}

\bibitem[{{Han} {et~al.}(2022{\natexlab{a}}){Han}, {Naidu}, {Conroy}, {Bonaca},
  {Zaritsky}, {Caldwell}, {Cargile}, {Johnson}, {Chandra}, {Speagle}, {Ting},
  \& {Woody}}]{han2022}
{Han}, J.~J., {Naidu}, R.~P., {Conroy}, C., {et~al.} 2022{\natexlab{a}}, \apj,
  934, 14, \dodoi{10.3847/1538-4357/ac795f}

\bibitem[{{Han} {et~al.}(2022{\natexlab{b}}){Han}, {Conroy}, {Johnson},
  {Speagle}, {Bonaca}, {Chandra}, {Naidu}, {Ting}, {Woody}, \&
  {Zaritsky}}]{han2022b}
{Han}, J.~J., {Conroy}, C., {Johnson}, B.~D., {et~al.} 2022{\natexlab{b}},
  arXiv e-prints, arXiv:2208.04327.
\newblock \doarXiv{2208.04327}

\bibitem[{Harris {et~al.}(2020)Harris, Millman, van~der Walt, Gommers,
  Virtanen, Cournapeau, Wieser, Taylor, Berg, Smith, Kern, Picus, Hoyer, van
  Kerkwijk, Brett, Haldane, del R{\'{i}}o, Wiebe, Peterson,
  G{\'{e}}rard-Marchant, Sheppard, Reddy, Weckesser, Abbasi, Gohlke, \&
  Oliphant}]{numpy}
Harris, C.~R., Millman, K.~J., van~der Walt, S.~J., {et~al.} 2020, Nature, 585,
  357, \dodoi{10.1038/s41586-020-2649-2}

\bibitem[{{Helmi}(2004)}]{helmi2004}
{Helmi}, A. 2004, \apjl, 610, L97, \dodoi{10.1086/423340}

\bibitem[{{Helmi} {et~al.}(2018){Helmi}, {Babusiaux}, {Koppelman}, {Massari},
  {Veljanoski}, \& {Brown}}]{Helmi2018}
{Helmi}, A., {Babusiaux}, C., {Koppelman}, H.~H., {et~al.} 2018, \nat, 563, 85,
  \dodoi{10.1038/s41586-018-0625-x}

\bibitem[{Helmi \& White(1999)}]{10.1046/j.1365-8711.1999.02616.x}
Helmi, A., \& White, S. D.~M. 1999, Monthly Notices of the Royal Astronomical
  Society, 307, 495, \dodoi{10.1046/j.1365-8711.1999.02616.x}

\bibitem[{{Hopkins}(2015)}]{Hopkins2015}
{Hopkins}, P.~F. 2015, \mnras, 450, 53

\bibitem[{{Hopkins} {et~al.}(2018){Hopkins}, {Wetzel}, {Kere{\v{s}}},
  {Faucher-Gigu{\`e}re}, {Quataert}, {Boylan-Kolchin}, {Murray}, {Hayward},
  {Garrison-Kimmel}, {Hummels}, {Feldmann}, {Torrey}, {Ma},
  {Angl{\'e}s-Alc{\'a}zar}, {Su}, {Orr}, {Schmitz}, {Escala}, {Sanderson},
  {Grudi{\'c}}, {Hafen}, {Kim}, {Fitts}, {Bullock}, {Wheeler}, {Chan},
  {Elbert}, \& {Narayanan}}]{2018MNRAS.480..800H}
{Hopkins}, P.~F., {Wetzel}, A., {Kere{\v{s}}}, D., {et~al.} 2018, \mnras, 480,
  800, \dodoi{10.1093/mnras/sty1690}

\bibitem[{{Hunter}(2007)}]{matplotlib}
{Hunter}, J.~D. 2007, Computing in Science and Engineering, 9, 90,
  \dodoi{10.1109/MCSE.2007.55}

\bibitem[{{Ibata} {et~al.}(1994){Ibata}, {Gilmore}, \& {Irwin}}]{ibata1994}
{Ibata}, R.~A., {Gilmore}, G., \& {Irwin}, M.~J. 1994, \nat, 370, 194,
  \dodoi{10.1038/370194a0}

\bibitem[{{Ivezi{\'c}} {et~al.}(2019){Ivezi{\'c}}, {Kahn}, {Tyson}, {Abel},
  {Acosta}, {Allsman}, {Alonso}, {AlSayyad}, {Anderson}, {Andrew}, {Angel},
  {Angeli}, {Ansari}, {Antilogus}, {Araujo}, {Armstrong}, {Arndt}, {Astier},
  {Aubourg}, {Auza}, {Axelrod}, {Bard}, {Barr}, {Barrau}, {Bartlett}, {Bauer},
  {Bauman}, {Baumont}, {Bechtol}, {Bechtol}, {Becker}, {Becla}, {Beldica},
  {Bellavia}, {Bianco}, {Biswas}, {Blanc}, {Blazek}, {Blandford}, {Bloom},
  {Bogart}, {Bond}, {Booth}, {Borgland}, {Borne}, {Bosch}, {Boutigny},
  {Brackett}, {Bradshaw}, {Brandt}, {Brown}, {Bullock}, {Burchat}, {Burke},
  {Cagnoli}, {Calabrese}, {Callahan}, {Callen}, {Carlin}, {Carlson},
  {Chandrasekharan}, {Charles-Emerson}, {Chesley}, {Cheu}, {Chiang}, {Chiang},
  {Chirino}, {Chow}, {Ciardi}, {Claver}, {Cohen-Tanugi}, {Cockrum}, {Coles},
  {Connolly}, {Cook}, {Cooray}, {Covey}, {Cribbs}, {Cui}, {Cutri}, {Daly},
  {Daniel}, {Daruich}, {Daubard}, {Daues}, {Dawson}, {Delgado}, {Dellapenna},
  {de Peyster}, {de Val-Borro}, {Digel}, {Doherty}, {Dubois},
  {Dubois-Felsmann}, {Durech}, {Economou}, {Eifler}, {Eracleous}, {Emmons},
  {Fausti Neto}, {Ferguson}, {Figueroa}, {Fisher-Levine}, {Focke}, {Foss},
  {Frank}, {Freemon}, {Gangler}, {Gawiser}, {Geary}, {Gee}, {Geha}, {Gessner},
  {Gibson}, {Gilmore}, {Glanzman}, {Glick}, {Goldina}, {Goldstein}, {Goodenow},
  {Graham}, {Gressler}, {Gris}, {Guy}, {Guyonnet}, {Haller}, {Harris},
  {Hascall}, {Haupt}, {Hernandez}, {Herrmann}, {Hileman}, {Hoblitt}, {Hodgson},
  {Hogan}, {Howard}, {Huang}, {Huffer}, {Ingraham}, {Innes}, {Jacoby}, {Jain},
  {Jammes}, {Jee}, {Jenness}, {Jernigan}, {Jevremovi{\'c}}, {Johns}, {Johnson},
  {Johnson}, {Jones}, {Juramy-Gilles}, {Juri{\'c}}, {Kalirai}, {Kallivayalil},
  {Kalmbach}, {Kantor}, {Karst}, {Kasliwal}, {Kelly}, {Kessler}, {Kinnison},
  {Kirkby}, {Knox}, {Kotov}, {Krabbendam}, {Krughoff}, {Kub{\'a}nek},
  {Kuczewski}, {Kulkarni}, {Ku}, {Kurita}, {Lage}, {Lambert}, {Lange},
  {Langton}, {Le Guillou}, {Levine}, {Liang}, {Lim}, {Lintott}, {Long},
  {Lopez}, {Lotz}, {Lupton}, {Lust}, {MacArthur}, {Mahabal}, {Mandelbaum},
  {Markiewicz}, {Marsh}, {Marshall}, {Marshall}, {May}, {McKercher}, {McQueen},
  {Meyers}, {Migliore}, {Miller}, {Mills}, {Miraval}, {Moeyens}, {Moolekamp},
  {Monet}, {Moniez}, {Monkewitz}, {Montgomery}, {Morrison}, {Mueller},
  {Muller}, {Mu{\~n}oz Arancibia}, {Neill}, {Newbry}, {Nief}, {Nomerotski},
  {Nordby}, {O'Connor}, {Oliver}, {Olivier}, {Olsen}, {O'Mullane}, {Ortiz},
  {Osier}, {Owen}, {Pain}, {Palecek}, {Parejko}, {Parsons}, {Pease},
  {Peterson}, {Peterson}, {Petravick}, {Libby Petrick}, {Petry},
  {Pierfederici}, {Pietrowicz}, {Pike}, {Pinto}, {Plante}, {Plate}, {Plutchak},
  {Price}, {Prouza}, {Radeka}, {Rajagopal}, {Rasmussen}, {Regnault}, {Reil},
  {Reiss}, {Reuter}, {Ridgway}, {Riot}, {Ritz}, {Robinson}, {Roby}, {Roodman},
  {Rosing}, {Roucelle}, {Rumore}, {Russo}, {Saha}, {Sassolas}, {Schalk},
  {Schellart}, {Schindler}, {Schmidt}, {Schneider}, {Schneider}, {Schoening},
  {Schumacher}, {Schwamb}, {Sebag}, {Selvy}, {Sembroski}, {Seppala}, {Serio},
  {Serrano}, {Shaw}, {Shipsey}, {Sick}, {Silvestri}, {Slater}, {Smith},
  {Smith}, {Sobhani}, {Soldahl}, {Storrie-Lombardi}, {Stover}, {Strauss},
  {Street}, {Stubbs}, {Sullivan}, {Sweeney}, {Swinbank}, {Szalay}, {Takacs},
  {Tether}, {Thaler}, {Thayer}, {Thomas}, {Thornton}, {Thukral}, {Tice},
  {Trilling}, {Turri}, {Van Berg}, {Vanden Berk}, {Vetter}, {Virieux},
  {Vucina}, {Wahl}, {Walkowicz}, {Walsh}, {Walter}, {Wang}, {Wang}, {Warner},
  {Wiecha}, {Willman}, {Winters}, {Wittman}, {Wolff}, {Wood-Vasey}, {Wu},
  {Xin}, {Yoachim}, \& {Zhan}}]{lsst}
{Ivezi{\'c}}, {\v{Z}}., {Kahn}, S.~M., {Tyson}, J.~A., {et~al.} 2019, \apj,
  873, 111, \dodoi{10.3847/1538-4357/ab042c}

\bibitem[{{Johnston} {et~al.}(2005){Johnston}, {Law}, \&
  {Majewski}}]{johnston2005}
{Johnston}, K.~V., {Law}, D.~R., \& {Majewski}, S.~R. 2005, \apj, 619, 800,
  \dodoi{10.1086/426777}

\bibitem[{{Kallivayalil} {et~al.}(2013){Kallivayalil}, {van der Marel},
  {Besla}, {Anderson}, \& {Alcock}}]{kallivayalil2013}
{Kallivayalil}, N., {van der Marel}, R.~P., {Besla}, G., {Anderson}, J., \&
  {Alcock}, C. 2013, \apj, 764, 161, \dodoi{10.1088/0004-637X/764/2/161}

\bibitem[{{Kollmeier} {et~al.}(2017){Kollmeier}, {Zasowski}, {Rix}, {Johns},
  {Anderson}, {Drory}, {Johnson}, {Pogge}, {Bird}, {Blanc}, {Brownstein},
  {Crane}, {De Lee}, {Klaene}, {Kreckel}, {MacDonald}, {Merloni}, {Ness},
  {O'Brien}, {Sanchez-Gallego}, {Sayres}, {Shen}, {Thakar}, {Tkachenko},
  {Aerts}, {Blanton}, {Eisenstein}, {Holtzman}, {Maoz}, {Nandra}, {Rockosi},
  {Weinberg}, {Bovy}, {Casey}, {Chaname}, {Clerc}, {Conroy}, {Eracleous},
  {G{\"a}nsicke}, {Hekker}, {Horne}, {Kauffmann}, {McQuinn}, {Pellegrini},
  {Schinnerer}, {Schlafly}, {Schwope}, {Seibert}, {Teske}, \& {van
  Saders}}]{sdss-v}
{Kollmeier}, J.~A., {Zasowski}, G., {Rix}, H.-W., {et~al.} 2017, arXiv
  e-prints, arXiv:1711.03234.
\newblock \doarXiv{1711.03234}

\bibitem[{{Laporte} {et~al.}(2018){Laporte}, {G{\'o}mez}, {Besla}, {Johnston},
  \& {Garavito-Camargo}}]{laporte2018}
{Laporte}, C. F.~P., {G{\'o}mez}, F.~A., {Besla}, G., {Johnston}, K.~V., \&
  {Garavito-Camargo}, N. 2018, \mnras, 473, 1218, \dodoi{10.1093/mnras/stx2146}

\bibitem[{{Law} {et~al.}(2005){Law}, {Johnston}, \& {Majewski}}]{law2005}
{Law}, D.~R., {Johnston}, K.~V., \& {Majewski}, S.~R. 2005, \apj, 619, 807,
  \dodoi{10.1086/426779}

\bibitem[{{Law} \& {Majewski}(2010)}]{law2010}
{Law}, D.~R., \& {Majewski}, S.~R. 2010, \apj, 714, 229,
  \dodoi{10.1088/0004-637X/714/1/229}

\bibitem[{{Leitherer} {et~al.}(1999){Leitherer}, {Schaerer}, {Goldader},
  {Delgado}, {Robert}, {Kune}, {de Mello}, {Devost}, \&
  {Heckman}}]{Leitherer1999}
{Leitherer}, C., {Schaerer}, D., {Goldader}, J.~D., {et~al.} 1999, \apjs, 123,
  3

\bibitem[{{Li} {et~al.}(2021){Li}, {Koposov}, {Erkal}, {Ji}, {Shipp}, {Pace},
  {Hilmi}, {Kuehn}, {Lewis}, {Mackey}, {Simpson}, {Wan}, {Zucker},
  {Bland-Hawthorn}, {Cullinane}, {Da Costa}, {Drlica-Wagner}, {Hattori},
  {Martell}, {Sharma}, \& {S5 Collaboration}}]{li2021}
{Li}, T.~S., {Koposov}, S.~E., {Erkal}, D., {et~al.} 2021, \apj, 911, 149,
  \dodoi{10.3847/1538-4357/abeb18}

\bibitem[{{Li} {et~al.}(2022){Li}, {Ji}, {Pace}, {Erkal}, {Koposov}, {Shipp},
  {Da Costa}, {Cullinane}, {Kuehn}, {Lewis}, {Mackey}, {Simpson}, {Zucker},
  {Ferguson}, {Martell}, {Bland-Hawthorn}, {Balbinot}, {Tavangar},
  {Drlica-Wagner}, {De Silva}, \& {Simon}}]{li2022}
{Li}, T.~S., {Ji}, A.~P., {Pace}, A.~B., {et~al.} 2022, \apj, 928, 30,
  \dodoi{10.3847/1538-4357/ac46d3}

\bibitem[{{Ma} {et~al.}(2017){Ma}, {Hopkins}, {Wetzel}, {Kirby},
  {Angl{\'e}s-Alc{\'a}zar}, {Faucher-Gigu{\`e}re}, {Kere{\v s}}, \&
  {Quataert}}]{Ma2017}
{Ma}, X., {Hopkins}, P.~F., {Wetzel}, A.~R., {et~al.} 2017, \mnras, 467, 2430

\bibitem[{{Majewski} {et~al.}(2003){Majewski}, {Skrutskie}, {Weinberg}, \&
  {Ostheimer}}]{Majewski2003}
{Majewski}, S.~R., {Skrutskie}, M.~F., {Weinberg}, M.~D., \& {Ostheimer}, J.~C.
  2003, \apj, 599, 1082, \dodoi{10.1086/379504}

\bibitem[{{Majewski} {et~al.}(2017){Majewski}, {Schiavon}, {Frinchaboy},
  {Allende Prieto}, {Barkhouser}, {Bizyaev}, {Blank}, {Brunner}, {Burton},
  {Carrera}, {Chojnowski}, {Cunha}, {Epstein}, {Fitzgerald}, {Garc{\'\i}a
  P{\'e}rez}, {Hearty}, {Henderson}, {Holtzman}, {Johnson}, {Lam}, {Lawler},
  {Maseman}, {M{\'e}sz{\'a}ros}, {Nelson}, {Nguyen}, {Nidever}, {Pinsonneault},
  {Shetrone}, {Smee}, {Smith}, {Stolberg}, {Skrutskie}, {Walker}, {Wilson},
  {Zasowski}, {Anders}, {Basu}, {Beland}, {Blanton}, {Bovy}, {Brownstein},
  {Carlberg}, {Chaplin}, {Chiappini}, {Eisenstein}, {Elsworth}, {Feuillet},
  {Fleming}, {Galbraith-Frew}, {Garc{\'\i}a}, {Garc{\'\i}a-Hern{\'a}ndez},
  {Gillespie}, {Girardi}, {Gunn}, {Hasselquist}, {Hayden}, {Hekker}, {Ivans},
  {Kinemuchi}, {Klaene}, {Mahadevan}, {Mathur}, {Mosser}, {Muna}, {Munn},
  {Nichol}, {O'Connell}, {Parejko}, {Robin}, {Rocha-Pinto}, {Schultheis},
  {Serenelli}, {Shane}, {Silva Aguirre}, {Sobeck}, {Thompson}, {Troup},
  {Weinberg}, \& {Zamora}}]{apogee}
{Majewski}, S.~R., {Schiavon}, R.~P., {Frinchaboy}, P.~M., {et~al.} 2017, \aj,
  154, 94, \dodoi{10.3847/1538-3881/aa784d}

\bibitem[{{Mateu}(2022)}]{galstream}
{Mateu}, C. 2022, arXiv e-prints, arXiv:2204.10326.
\newblock \doarXiv{2204.10326}

\bibitem[{{Miyamoto} \& {Nagai}(1975)}]{1975PASJ...27..533M}
{Miyamoto}, M., \& {Nagai}, R. 1975, \pasj, 27, 533

\bibitem[{{Newberg} {et~al.}(2002){Newberg}, {Yanny}, {Rockosi}, {Grebel},
  {Rix}, {Brinkmann}, {Csabai}, {Hennessy}, {Hindsley}, {Ibata}, {Ivezi{\'c}},
  {Lamb}, {Nash}, {Odenkirchen}, {Rave}, {Schneider}, {Smith}, {Stolte}, \&
  {York}}]{Newberg2002}
{Newberg}, H.~J., {Yanny}, B., {Rockosi}, C., {et~al.} 2002, \apj, 569, 245,
  \dodoi{10.1086/338983}

\bibitem[{{Niederste-Ostholt} {et~al.}(2010){Niederste-Ostholt}, {Belokurov},
  {Evans}, \& {Pe{\~n}arrubia}}]{2010ApJ...712..516N}
{Niederste-Ostholt}, M., {Belokurov}, V., {Evans}, N.~W., \& {Pe{\~n}arrubia},
  J. 2010, \apj, 712, 516, \dodoi{10.1088/0004-637X/712/1/516}

\bibitem[{{Panithanpaisal} {et~al.}(2021){Panithanpaisal}, {Sanderson},
  {Wetzel}, {Cunningham}, {Bailin}, \&
  {Faucher-Gigu{\`e}re}}]{Panithanpaisal2021}
{Panithanpaisal}, N., {Sanderson}, R.~E., {Wetzel}, A., {et~al.} 2021, \apj,
  920, 10, \dodoi{10.3847/1538-4357/ac1109}

\bibitem[{{Pe{\~n}arrubia} {et~al.}(2011){Pe{\~n}arrubia}, {Zucker}, {Irwin},
  {Hyde}, {Lane}, {Lewis}, {Gilmore}, {Evans}, \&
  {Belokurov}}]{2011ApJ...727L...2P}
{Pe{\~n}arrubia}, J., {Zucker}, D.~B., {Irwin}, M.~J., {et~al.} 2011, \apjl,
  727, L2, \dodoi{10.1088/2041-8205/727/1/L2}

\bibitem[{{Pearson} {et~al.}(2022){Pearson}, {Clark}, {Demirjian}, {Johnston},
  {Ness}, {Starkenburg}, {Williams}, \& {Ibata}}]{2022ApJ...926..166P}
{Pearson}, S., {Clark}, S.~E., {Demirjian}, A.~J., {et~al.} 2022, \apj, 926,
  166, \dodoi{10.3847/1538-4357/ac4496}

\bibitem[{{Pearson} {et~al.}(2017){Pearson}, {Price-Whelan}, \&
  {Johnston}}]{pearon2017}
{Pearson}, S., {Price-Whelan}, A.~M., \& {Johnston}, K.~V. 2017, Nature
  Astronomy, 1, 633, \dodoi{10.1038/s41550-017-0220-3}

\bibitem[{Perez \& Granger(2007)}]{ipython}
Perez, F., \& Granger, B.~E. 2007, Computing in Science \& Engineering, 9, 21,
  \dodoi{10.1109/MCSE.2007.53}

\bibitem[{{Petersen} \& {Pe{\~n}arrubia}(2021)}]{petersen2021}
{Petersen}, M.~S., \& {Pe{\~n}arrubia}, J. 2021, Nature Astronomy, 5, 251,
  \dodoi{10.1038/s41550-020-01254-3}

\bibitem[{{Reino} {et~al.}(2021){Reino}, {Rossi}, {Sanderson}, {Sellentin},
  {Helmi}, {Koppelman}, \& {Sharma}}]{reino2021}
{Reino}, S., {Rossi}, E.~M., {Sanderson}, R.~E., {et~al.} 2021, \mnras, 502,
  4170, \dodoi{10.1093/mnras/stab304}

\bibitem[{{Sanders} \& {Binney}(2013{\natexlab{a}})}]{sanders2013b}
{Sanders}, J.~L., \& {Binney}, J. 2013{\natexlab{a}}, \mnras, 433, 1826,
  \dodoi{10.1093/mnras/stt816}

\bibitem[{{Sanders} \& {Binney}(2013{\natexlab{b}})}]{sanders2013a}
---. 2013{\natexlab{b}}, \mnras, 433, 1813, \dodoi{10.1093/mnras/stt806}

\bibitem[{{Sanderson} {et~al.}(2017){Sanderson}, {Hartke}, \&
  {Helmi}}]{sanderson2017}
{Sanderson}, R.~E., {Hartke}, J., \& {Helmi}, A. 2017, \apj, 836, 234,
  \dodoi{10.3847/1538-4357/aa5eb4}

\bibitem[{{Sanderson} {et~al.}(2020){Sanderson}, {Wetzel}, {Loebman}, {Sharma},
  {Hopkins}, {Garrison-Kimmel}, {Faucher-Gigu{\`e}re}, {Kere{\v{s}}}, \&
  {Quataert}}]{2020ApJS..246....6S}
{Sanderson}, R.~E., {Wetzel}, A., {Loebman}, S., {et~al.} 2020, \apjs, 246, 6,
  \dodoi{10.3847/1538-4365/ab5b9d}

\bibitem[{{Shao} {et~al.}(2021){Shao}, {Cautun}, {Deason}, \&
  {Frenk}}]{shao2020}
{Shao}, S., {Cautun}, M., {Deason}, A., \& {Frenk}, C.~S. 2021, \mnras, 504,
  6033, \dodoi{10.1093/mnras/staa3883}

\bibitem[{{Shipp} {et~al.}(2018){Shipp}, {Drlica-Wagner}, {Balbinot},
  {Ferguson}, {Erkal}, {Li}, {Bechtol}, {Belokurov}, {Buncher}, {Carollo},
  {Carrasco Kind}, {Kuehn}, {Marshall}, {Pace}, {Rykoff}, {Sevilla-Noarbe},
  {Sheldon}, {Strigari}, {Vivas}, {Yanny}, {Zenteno}, {Abbott}, {Abdalla},
  {Allam}, {Avila}, {Bertin}, {Brooks}, {Burke}, {Carretero}, {Castander},
  {Cawthon}, {Crocce}, {Cunha}, {D'Andrea}, {da Costa}, {Davis}, {De Vicente},
  {Desai}, {Diehl}, {Doel}, {Evrard}, {Flaugher}, {Fosalba}, {Frieman},
  {Garc{\'\i}a-Bellido}, {Gaztanaga}, {Gerdes}, {Gruen}, {Gruendl}, {Gschwend},
  {Gutierrez}, {Hartley}, {Honscheid}, {Hoyle}, {James}, {Johnson}, {Krause},
  {Kuropatkin}, {Lahav}, {Lin}, {Maia}, {March}, {Martini}, {Menanteau},
  {Miller}, {Miquel}, {Nichol}, {Plazas}, {Romer}, {Sako}, {Sanchez},
  {Santiago}, {Scarpine}, {Schindler}, {Schubnell}, {Smith}, {Smith},
  {Sobreira}, {Suchyta}, {Swanson}, {Tarle}, {Thomas}, {Tucker}, {Walker},
  {Wechsler}, \& {DES Collaboration}}]{Shipp2018}
{Shipp}, N., {Drlica-Wagner}, A., {Balbinot}, E., {et~al.} 2018, \apj, 862,
  114, \dodoi{10.3847/1538-4357/aacdab}

\bibitem[{{Shipp} {et~al.}(2019){Shipp}, {Li}, {Pace}, {Erkal},
  {Drlica-Wagner}, {Yanny}, {Belokurov}, {Wester}, {Koposov}, {Kuehn}, {Lewis},
  {Simpson}, {Wan}, {Zucker}, {Martell}, {Wang}, \& {S5
  Collaboration}}]{shipp2019}
{Shipp}, N., {Li}, T.~S., {Pace}, A.~B., {et~al.} 2019, \apj, 885, 3,
  \dodoi{10.3847/1538-4357/ab44bf}

\bibitem[{{Shipp} {et~al.}(2021){Shipp}, {Erkal}, {Drlica-Wagner}, {Li},
  {Pace}, {Koposov}, {Cullinane}, {Da Costa}, {Ji}, {Kuehn}, {Lewis}, {Mackey},
  {Simpson}, {Wan}, {Zucker}, {Bland-Hawthorn}, {Ferguson}, {Lilleengen}, \&
  {Lilleengen}}]{shipp2021}
{Shipp}, N., {Erkal}, D., {Drlica-Wagner}, A., {et~al.} 2021, \apj, 923, 149,
  \dodoi{10.3847/1538-4357/ac2e93}

\bibitem[{{Skrutskie} {et~al.}(2006){Skrutskie}, {Cutri}, {Stiening},
  {Weinberg}, {Schneider}, {Carpenter}, {Beichman}, {Capps}, {Chester},
  {Elias}, {Huchra}, {Liebert}, {Lonsdale}, {Monet}, {Price}, {Seitzer},
  {Jarrett}, {Kirkpatrick}, {Gizis}, {Howard}, {Evans}, {Fowler}, {Fullmer},
  {Hurt}, {Light}, {Kopan}, {Marsh}, {McCallon}, {Tam}, {Van Dyk}, \&
  {Wheelock}}]{2mass}
{Skrutskie}, M.~F., {Cutri}, R.~M., {Stiening}, R., {et~al.} 2006, \aj, 131,
  1163, \dodoi{10.1086/498708}

\bibitem[{{Spergel} {et~al.}(2013){Spergel}, {Gehrels}, {Breckinridge},
  {Donahue}, {Dressler}, {Gaudi}, {Greene}, {Guyon}, {Hirata}, {Kalirai},
  {Kasdin}, {Moos}, {Perlmutter}, {Postman}, {Rauscher}, {Rhodes}, {Wang},
  {Weinberg}, {Centrella}, {Traub}, {Baltay}, {Colbert}, {Bennett},
  {Kiessling}, {Macintosh}, {Merten}, {Mortonson}, {Penny}, {Rozo},
  {Savransky}, {Stapelfeldt}, {Zu}, {Baker}, {Cheng}, {Content}, {Dooley},
  {Foote}, {Goullioud}, {Grady}, {Jackson}, {Kruk}, {Levine}, {Melton},
  {Peddie}, {Ruffa}, \& {Shaklan}}]{wfirst}
{Spergel}, D., {Gehrels}, N., {Breckinridge}, J., {et~al.} 2013, arXiv
  e-prints, arXiv:1305.5422.
\newblock \doarXiv{1305.5422}

\bibitem[{Tremaine(1999)}]{10.1046/j.1365-8711.1999.02690.x}
Tremaine, S. 1999, Monthly Notices of the Royal Astronomical Society, 307, 877,
  \dodoi{10.1046/j.1365-8711.1999.02690.x}

\bibitem[{{Vasiliev}(2019)}]{2019MNRAS.482.1525V}
{Vasiliev}, E. 2019, \mnras, 482, 1525, \dodoi{10.1093/mnras/sty2672}

\bibitem[{{Vasiliev} \& {Belokurov}(2020)}]{vasiliev2020}
{Vasiliev}, E., \& {Belokurov}, V. 2020, \mnras, 497, 4162,
  \dodoi{10.1093/mnras/staa2114}

\bibitem[{{Vasiliev} {et~al.}(2021){Vasiliev}, {Belokurov}, \&
  {Erkal}}]{vasiliev2021}
{Vasiliev}, E., {Belokurov}, V., \& {Erkal}, D. 2021, \mnras, 501, 2279,
  \dodoi{10.1093/mnras/staa3673}

\bibitem[{Virtanen {et~al.}(2020)Virtanen, Gommers, Oliphant, Haberland, Reddy,
  Cournapeau, Burovski, Peterson, Weckesser, Bright, {van der Walt}, Brett,
  Wilson, Millman, Mayorov, Nelson, Jones, Kern, Larson, Carey, Polat, Feng,
  Moore, {VanderPlas}, Laxalde, Perktold, Cimrman, Henriksen, Quintero, Harris,
  Archibald, Ribeiro, Pedregosa, {van Mulbregt}, \& {SciPy 1.0
  Contributors}}]{scipy}
Virtanen, P., Gommers, R., Oliphant, T.~E., {et~al.} 2020, Nature Methods, 17,
  261, \dodoi{10.1038/s41592-019-0686-2}

\bibitem[{{W}es {M}c{K}inney(2010)}]{pandas2}
{W}es {M}c{K}inney. 2010, in {P}roceedings of the 9th {P}ython in {S}cience
  {C}onference, ed. {S}t\'efan van~der {W}alt \& {J}arrod {M}illman, 56 -- 61,
  \dodoi{10.25080/Majora-92bf1922-00a}

\bibitem[{{Wetzel} \&
  {Garrison-Kimmel}(2020{\natexlab{a}})}]{2020ascl.soft02014W}
{Wetzel}, A., \& {Garrison-Kimmel}, S. 2020{\natexlab{a}}, {HaloAnalysis: Read
  and analyze halo catalogs and merger trees}.
\newblock \doeprint{2002.014}

\bibitem[{{Wetzel} \&
  {Garrison-Kimmel}(2020{\natexlab{b}})}]{2020ascl.soft02015W}
---. 2020{\natexlab{b}}, {GizmoAnalysis: Read and analyze Gizmo simulations}.
\newblock \doeprint{2002.015}

\bibitem[{{Wetzel} {et~al.}(2022){Wetzel}, {Hayward}, {Sanderson}, {Ma},
  {Angles-Alcazar}, {Feldmann}, {Chan}, {El-Badry}, {Wheeler},
  {Garrison-Kimmel}, {Nikakhtar}, {Panithanpaisal}, {Arora}, {Gurvich},
  {Samuel}, {Sameie}, {Pandya}, {Hafen}, {Hummels}, {Loebman},
  {Boylan-Kolchin}, {Bullock}, {Faucher-Giguere}, {Keres}, {Quataert}, \&
  {Hopkins}}]{2022arXiv220206969W}
{Wetzel}, A., {Hayward}, C.~C., {Sanderson}, R.~E., {et~al.} 2022, arXiv
  e-prints, arXiv:2202.06969.
\newblock \doarXiv{2202.06969}

\bibitem[{{Wetzel} {et~al.}(2016){Wetzel}, {Hopkins}, {Kim},
  {Faucher-Gigu{\`e}re}, {Kere{\v s}}, \& {Quataert}}]{Wetzel2016}
{Wetzel}, A.~R., {Hopkins}, P.~F., {Kim}, J.-h., {et~al.} 2016, \apjl, 827, L23

\bibitem[{{Yavetz} {et~al.}(2021){Yavetz}, {Johnston}, {Pearson},
  {Price-Whelan}, \& {Weinberg}}]{tomer}
{Yavetz}, T.~D., {Johnston}, K.~V., {Pearson}, S., {Price-Whelan}, A.~M., \&
  {Weinberg}, M.~D. 2021, \mnras, 501, 1791, \dodoi{10.1093/mnras/staa3687}

\bibitem[{Zonca {et~al.}(2019)Zonca, Singer, Lenz, Reinecke, Rosset, Hivon, \&
  Gorski}]{Zonca2019}
Zonca, A., Singer, L., Lenz, D., {et~al.} 2019, Journal of Open Source
  Software, 4, 1298, \dodoi{10.21105/joss.01298}

\end{thebibliography}
\end{document}